\documentclass{aa}
\usepackage{comment}
\usepackage{amsmath}
\usepackage{rotating}
\usepackage{multirow}
\usepackage{multicol}
\usepackage{hyperref}
\usepackage{multirow}
\usepackage{array}
\usepackage{longtable}
\usepackage{tablefootnote}
\usepackage{ulem}
\hypersetup{
    colorlinks=true,            
    linkcolor=black,             
    citecolor=blue,              
    urlcolor=blue               
}
\usepackage{graphicx}
\usepackage{txfonts}
\usepackage{hyperref}
\usepackage{academicons}
\usepackage{xcolor}
\usepackage{natbib}
\bibpunct{(}{)}{;}{a}{}{,} 
\begin{document} 

  \title{Revisiting the exoplanet radius valley with host stars from SWEET-Cat}

   \author{J. Kamulali
          \inst{1,2,3}\href{https://orcid.org/0009-0005-5492-8482}{\includegraphics[scale=0.07]{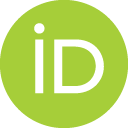}},
          V. Adibekyan\inst{3,4}\href{https://orcid.org/0000-0002-0601-6199}{\includegraphics[scale=0.07]{figures/ORCID_logo.png}},
          B. Nsamba\inst{1,2,5}\href{https://orcid.org/0000-0002-4647-2068}{\includegraphics[scale=0.07]{figures/ORCID_logo.png}},
          S.G. Sousa\inst{3,4}\href{https://orcid.org/0000-0001-9047-2965}{\includegraphics[scale=0.07]{figures/ORCID_logo.png}}, 
          T. L. Campante\inst{3,4}\href{https://orcid.org/0000-0002-4588-5389}{\includegraphics[scale=0.07]{figures/ORCID_logo.png}},
           A. Weiss\inst{2}\href{https://orcid.org/0000-0002-3843-1653}{\includegraphics[scale=0.07]{figures/ORCID_logo.png}},
           B. Kabugho\inst{1}\href{https://orcid.org/0009-0009-4890-9292}{\includegraphics[scale=0.07]{figures/ORCID_logo.png}},
          N. Moedas\inst{6}\href{https://orcid.org/0000-0002-2087-6427}{\includegraphics[scale=0.07]{figures/ORCID_logo.png}},
          N. C. Santos\inst{3,4}\href{https://orcid.org/0000-0003-4422-2919}{\includegraphics[scale=0.07]{figures/ORCID_logo.png}},
          \and
          O. Trust\inst{1,5,7}\href{https://orcid.org/0000-0001-5209-7041}{\includegraphics[scale=0.07]{figures/ORCID_logo.png}}   
        }
          
   \institute{Department of Physics, Faculty of Science, Kyambogo University, P.O. Box 1, Kyambogo, Uganda\\
              \email{kamulali@mpa-garching.mpg.de}
         \and
             Max-Planck-Institut f\"{u}r Astrophysik, Karl-Schwarzschild-Str. 1, D-85748 Garching, Germany
         \and
            Instituto de Astrof\'{\i}sica e Ci\^{e}ncias do Espa\c{c}o, Universidade do Porto, Rua das Estrelas, 4150-762 Porto, Portugal
         \and 
            Departamento de F\'{\i}sica e Astronomia, Faculdade de Ci\^{e}ncias da Universidade do Porto, Rua do Campo Alegre, s/n, 4169-007 Porto, Portugal
         \and
         Max-Planck-Institut f\"{u}r Sonnensystemforschung, Justus-von-Liebig-Weg 3, 37077 G\"{o}ttingen, Germany
         \and
            INAF-IAPS, Via del Fosso del Cavaliere 100, 00133 Roma, Italy
         \and 
            Department of Physics, Mbarara University of Science and Technology, P.O. Box 1410, Mbarara, Uganda
             }

   \date{\textbf{Received:}  06 October 2025  \textbf{Accepted:} 13 January 2026}
   \titlerunning{Radius valley}
   \authorrunning{Kamulali et al.}
 
  \abstract
   {The radius valley, a deficit in the number of planets with radii around 2 R$_\oplus$, was observed among exoplanets with sizes $\lesssim$ 5 R$_\oplus$ and orbital periods $<$ 100 days by NASA's $Kepler$ mission. This feature separates two distinct populations: super-Earths (rocky planets with radii $\lesssim$ 1.9 R$_\oplus$) and sub-Neptunes (planets with substantial volatile envelopes and radii $\gtrsim$ 2 R$_\oplus$). The valley has been proposed to stem from either planet formation conditions or evolutionary atmospheric loss processes. Disentangling these mechanisms has led to numerous studies of population-level trends, although the resulting interpretations remain sensitive to sample selection and the robustness of host-star parameters.} 
   {To re-examine the existence and depth of the radius valley, and how its location varies with orbital period, incident flux, stellar mass, and stellar age.}
   {We derived robust fundamental stellar parameters of 1,221 main-sequence stars (hosting 1,405 confirmed planets) from the SWEET-Cat database using a grid-based machine-learning tool (MAISTEP), which incorporates effective temperatures and metallicities from spectroscopy, as well as $Gaia$-based luminosities. Our analysis covers stars with effective temperatures between 4400 -- 7500 K (FGK spectral types) and estimated radii between 0.62 -- 2.75 R$_\odot$. We attained an average uncertainty of 2\% in stellar radius and 2\% in mass. Combining the updated stellar radii with planet-to-star radius ratios from the NASA Exoplanet Archive, we recomputed the planetary radii achieving an average uncertainty of 5\%.}
   {Our findings confirm a partially filled planet radius valley near 2 R$_\oplus$. The valley depends on orbital period, incident flux, and stellar mass, with slopes of $-0.12^{+0.02}_{-0.01}$, $0.10^{+0.02}_{-0.03}$, $0.19^{+0.09}_{-0.07}$, respectively. We also find a stronger mass-dependent trend in average sizes of sub-Neptunes than super-Earths of slopes $0.17^{+0.04}_{-0.04}$ and $0.11 ^{+0.05}_{-0.05}$, respectively. 
   With stellar age, the super-Earth/sub-Neptune number ratio increases from $0.51^{+0.11}_{-0.08}$ ($< 3$ Gyr) to $0.64^{+0.11}_{-0.11}$ ($\geq 3$ Gyr). In addition, the valley becomes shallower and shifts to larger radii, indicating age-dependent evolution in planet sizes. A four-dimensional (planet radius, orbital period, stellar mass, and stellar age) linear fit of the valley produces slopes in orbital period and stellar mass that are consistent with the results of the two-dimensional analyses, and a weaker slope of $0.07^{+0.03}_{-0.04}$ in stellar age.} 
   {The valley's shift and shallowing over gigayear timescales point to prolonged atmospheric loss, which is consistent with a core-powered mass-loss scenario. Our findings also highlight the importance of stellar age in the interpretation of exoplanet demographics and motivate improved age determinations, as expected from future missions like \textit{PLATO}.}
   \keywords{planetary systems / planets and satellites: formation / stars: fundamental parameters}

\maketitle

\section{Introduction}

Observations from NASA’s $Kepler$ mission revealed more than 4,000 exoplanets, most with sizes $\lesssim 5$ R$_\oplus$ and orbital periods $<$ 100 days \citep[][]{borucki2010kepler}. A notable feature of this population is a bimodal distribution in the planetary radii, with peaks near $\sim$1.5 and $\sim$2.4 R$_\oplus$ \citep[][]{fulton2017california}. The region in between these peaks is referred to as the "radius gap" or "radius valley." 

The radius valley was not apparent in the initial $Kepler$ data due to large uncertainties in the stellar radii, with typical median errors $\sim$ 25\% \citep[][]{huber2014revised}. Subsequent improvement in stellar characterisation from the California-Kepler Survey (CKS; \citealt{johnson2017california, petigura2017california}), reduced the uncertainties significantly. \citet{fulton2017california} reported a median uncertainty of 11\% in stellar radii for the CKS sample, translating to a 12\% uncertainty in planetary radii. This level of precision allowed for the first observational measurement of a valley among close-in planets, uncovering a key demographic feature in the $Kepler$ observations.

Following this detection, the question of whether the radius valley is truly devoid of planets or partially populated gained attention. In \citet{fulton2017california}, the reported planet radius uncertainties of $\sim$ 12\% in their selected CKS sample of 900 planets, limited to host stars with temperatures between 4700 K and 6500 K, were comparable to the valley’s width, making it difficult to determine whether the gap was intrinsically empty. Based on asteroseismic stellar radii with a median fractional uncertainty of 2.2\% and the corresponding planetary radii with a median uncertainty of 3.3\%, \citet{van2018asteroseismic} found the valley to be wider and emptier. However, their result may have been influenced by the relatively small sample size of 117 planets. In another study incorporating Gaia DR2 parallaxes, \citet{fulton2018california} improved stellar radius measurements to 3\%, achieving 5\% precision in planetary radii for $\sim$ 1000 planets, and found the valley to be populated, a conclusion further supported by their simulations. Later, \citet{petigura2020two} revisited the question and argued that many planets near the valley have poorly constrained radii due to significant uncertainties in their measured planet-star radius ratios, an issue common among planets transiting at high impact parameters ($b\gtrsim 0.8$), suggesting the gap may, in fact, be largely empty. Resolving the nature of the radius valley is pivotal in refining models of planetary formation and evolution developed to explain its existence.

To account for this feature in the exoplanet population, several mechanisms have been proposed. In the photoevaporation model, intense high-energy (XUV) radiation from the host star strips away the primordial hydrogen–helium envelopes of close-in, low-mass planets \citep[][]{owen2013kepler,lopez2013role}. This atmospheric loss naturally produces a bimodal planet-size distribution: rocky, high-density super-Earths below the valley, and low-density sub-Neptunes above it, which retain significant gaseous envelopes. 

Core-powered mass-loss mechanism has also been shown to reproduce the observed radius valley \citep[][]{ginzburg2016super,ginzburg2018core,gupta2019sculpting,gupta2020signatures}. In this scenario, atmospheric escape is driven by a planet’s internal luminosity, powered by the residual heat from formation, rather than by external stellar irradiation. Photoevaporation or core-powered mass loss reproduce the valley only if the stripped cores are assumed to be rocky, suggesting that most Earth–Neptune size $Kepler$ planets form inside the ice line from dry material. By contrast, \citet{venturini2020nature,venturini2020most} used a hybrid model to show that the sub-Neptune peak comes from water-rich planets migrating inward from beyond the ice line, whereas the super-Earth peak originates from in-situ rocky planets whose atmospheres were stripped by photoevaporation \citep[also see,][]{2024arXiv241116879B,burn2024radius,chakrabarty2024water}.

Other proposed explanations include atmospheric stripping through giant impacts, generating shock waves from core oscillations that propagate through the planetary envelope, potentially removing substantial portions of the atmosphere \citep[][]{liu2015giant,schlichting2015atmospheric,ogihara2020unified}, and late-time formation in gas-poor environments, where rocky super-Earths form after the dispersal of the protoplanetary gas disk and therefore never accumulate atmospheres \citep[][]{lee2014make,lee2016breeding,lopez2018formation,lee2019episodic}.

Among these mechanisms, photoevaporation and core-powered mass-loss have been shown to recover the observed dependence of the radius valley on orbital period (or incident flux) and stellar mass for planets orbiting solar-type stars \citep[e.g.,][]{van2018asteroseismic,wu2019mass,berger2020gaiab,cloutier2020evolution,gupta2020signatures,van2021masses,petigura2022california}. Efforts to unravel their individual contributions are ongoing. For example, \citet{rogers2021photoevaporation} examined the valley in three-dimensional space (radius, stellar mass, and incident flux) based on observational data from the California-Kepler and Gaia-Kepler surveys \citep[][]{fulton2017california,fulton2018california,berger2020gaiaa}. They found that the observational data distributions were consistent with both mechanisms, attributing the degeneracy primarily to the limited sample size. \citet{2023arXiv230200009B}, using an expanded dataset from the Gaia-Kepler-TESS Host Properties catalogue \citep[][]{2023arXiv230111338B}, reported three-dimensional trends more consistent with the predictions of core-powered mass loss than those of photoevaporation. Similarly, \citet{ho2023deep} reached the same conclusion using stellar ages from isochrone fitting.

Given that the radii of close-in planets with light atmospheres and/or water (volatiles) are expected to change over time, their variation with host star age has also been investigated in an attempt to distinguish between atmospheric loss mechanisms. In photoevaporation models, high-energy radiation drives atmospheric escape primarily within the first $\sim$100 Myr, coinciding with the peak XUV emission of Sun-like stars \citep[e.g.,][]{jackson2012coronal,tu2015extreme}. Core-powered mass loss, on the other hand, is driven by a planet’s internal cooling and operates continuously over gigayear timescales \citep[][]{ginzburg2016super, ginzburg2018core, gupta2019sculpting, gupta2020signatures}. Based on this difference, observational trends with system age have been used to discriminate between these mechanisms.

For example, \citet{berger2020gaiab} found that the number of super-Earths to sub-Neptunes increases with stellar age over gigayear timescales. Similarly, \citet{sandoval2021influence} reported an increase in the super-Earths to sub-Neptunes ratio with stellar age in the $\sim 1 -10$ Gyr range. In addition, \citet{david2021evolution} provided evidence that the precise location of the radius valley changes over gigayear timescales, with the valley appearing emptier below 3 Gyr due to a deficit of large super-Earths ($\sim 1.5 - 1.8$ R$_\oplus$) orbiting young stars. This absence makes the valley appear at smaller planet radii in younger systems than in older ones. These results could be interpreted as supporting core-powered mass loss, which predicts ongoing atmospheric erosion long after the early, high-radiation phase assumed in photoevaporation models. However, subsequent work by \citet{king2021euv} showed that extreme UV irradiation persists beyond 100 Myr. If both mechanisms operate on overlapping timescales, it could complicate efforts to identify a dominant process based solely on age. In addition to changes in occurrence rates, sub-Neptunes are expected to shrink over time. Yet, direct evidence for a decrease in sub-Neptunes remains inconclusive. Using isochrone-based stellar ages, \citet{petigura2022california} found no significant correlation between sub-Neptune sizes and age, possibly due to large uncertainties in age estimates. By comparison, \citet{chen2022planets}, employing kinematic ages derived from velocity dispersion in a LAMOST–Gaia–Kepler sample, reported a measurable decline in the radii of 238 sub-Neptunes with increasing age.

 These studies collectively highlight the importance of large, well-characterised samples and precise stellar parameters for interpreting exoplanet population trends. Using the Machine learning Algorithms for Inferring STEllar Parameters \citep[MAISTEP:][]{kamulali2025maistep}, which incorporates effective temperatures and metallicities from spectroscopy, as well as $Gaia$-based luminosities, we achieve a precision of 2\% in radius and 3\% in mass for main-sequence stars, comparable to asteroseismic measurements. Applying this approach to characterise a sample of 1,221 main-sequence stars (hosting 1,405 confirmed planets) from the SWEET-Cat database \citep[][]{santos2013sweet,sousa2021sweet,sousa2024sweet}, we infer robust stellar radius, mass, and age. With these updated parameters, we re-examine the existence and depth of the radius valley, and how its location varies with orbital period, incident flux, stellar mass, and stellar age.

  The rest of the paper is organised as follows: In Sect.\ref{method}, we describe the generation of the MAISTEP model, along with details of the host star and planet samples. Sect.\ref{gap} presents our results on the radius valley, including its dependence on orbital and stellar parameters. We conclude our findings in Sect.~\ref{conclusion}.

\section{Method}
\label{method}
\subsection{MAISTEP model}

The MAISTEP tool \citep[][]{kamulali2025maistep} employs a stacked ensemble of four tree-based algorithms i.e., random forest \citep[][]{breiman2001random}, extra trees \citep[][]{wehenkel2006ensembles}, extreme gradient boosting \citep[][]{chen2016xgboost}, and CatBoost \citep[][]{prokhorenkova2018catboost}. These algorithms are trained on a grid of stellar evolutionary tracks to infer stellar radius, mass, and age using effective temperature, metallicity, and luminosity. Each algorithm generates independent predictions, which are combined using a weighting criterion to produce the final parameter estimates. In this study, MAISTEP was trained on stellar evolutionary models from \cite{moedas2024characterisation}, specifically considering grid A that was computed without radiative acceleration. The evolutionary tracks run from the zero-age main sequence to the end of the main sequence, defined as the point along the track when the central hydrogen mass fraction drops below 0.001. Atomic diffusion (gravitation settling component only; \citealt{1994ApJ...421..828T}) was applied selectively to prevent excessive surface depletion of heavy elements based on the initial chemical composition (the extent of the convective envelope depends on the chemical composition). All main-sequence models with masses below 1.0 M$_{\odot}$ include diffusion. Models with masses higher than 1.4 M$_{\odot}$ exclude diffusion entirely. For intermediate-mass models, diffusion was applied conditionally, with its inclusion becoming rare as the model mass increases. Details of this conditioning are explored in \citet{moedas2022atomic,moedas2024characterisation}. We also note that no rotational mixing or mass loss was included in the models. In Table~\ref{global_physics}, we summarise the initial grid parameter ranges and adopted global physics.

\begin{table}[t]
\small
\caption{Initial grid parameter ranges and input physics. $\Delta$ refers to the step size in our Cartesian grid.}
    \centering
    \begin{tabular}{lc}
    \hline\hline
    Parameter & Range \\
    \hline
    Initial mass in M$_{\odot}$  & [0.7 -- 1.75], $\Delta M = 0.05$\\
    Initial metallicity in dex & [-0.4 -- 0.5], $\Delta$[Fe/H] = 0.05\\
    Initial helium & [0.24 -- 0.34], $\Delta$$Y_{\text{i}}$ = 0.01\\
    \hline
    \multicolumn{2}{c}{}\\
    \hline
        Input physics & Prescription \\
        \hline
        Nuclear reaction rates  & 1, 2, 3\\
        Solar element mixtures & 4\\
        OPAL equation of state & 5\\
        OPAL opacities  &  6, 7\\
        Atmosphere model & 8\\
        Atomic diffusion  & 9\\
        Core overshooting, $f_{\text{ov}} = 0.01$ & 10\\
        $\alpha_{\text{MLT}}= 1.71$ & 11\\
        \hline
    \end{tabular}

   \tablebib{ (1)~\citet{angulo1999compilation};  (2) \citet{imbriani2005s}; (3) \citet{kunz2002astrophysical}; (4) \citet{asplund2009chemical};  (5) \citet{rogers2002updated};  (6) \citet{iglesias1996updated};  (7) \citet{ferguson2005low}; (8) \citet{krishna1966profiles};  (9) \citet{1994ApJ...421..828T};  (10) \citet{herwig2000evolution};  (11) \citet{cox1968principles}.}
    \label{global_physics}
\end{table}

\subsection{Stellar and planetary sample}
\label{sample}
The SWEET-Cat\footnote{https://sweetcat.iastro.pt/} database (Stars With ExoplanETs catalogue) provides an updated list of parameters for stars hosting planets \citep[][]{santos2013sweet,sousa2021sweet,sousa2024sweet}. Currently, the catalogue is composed of 4,206 planet hosts ranging from the main sequence to more evolved phases. 
Of these, 1,191 stars (28\%) have spectroscopic effective temperature, metallicity, and surface gravity, determined homogeneously. The homogeneous host star sample would be ideal for this study; however, the available sample size is limited. To ensure a sufficiently large number of host stars, our analysis includes all main-sequence stars that fall within the bounds of our MAISTEP model training grid in effective temperature, metallicity, and surface gravity. Star selection is based on matching observed parameters to the model parameters of this grid. A star is included only if more than 100 grid points lie within $\pm$ 84 K in effective temperature, $\pm$ 0.05 dex in metallicity and $\pm0.05$ cm/s$^2$ in surface gravity. These tolerances reflect the mean parameter uncertainties and define a sample of 2,128 main-sequence host stars. They were used only for selection and do not represent the uncertainty limits for the sample. We use input parameter uncertainties directly from SWEET-Cat. However, only a subset of these stars host planets with measured planet-to-star radius ratios, which are required to compute planetary radii. 

To obtain the planet-to-star radius ratios, we use the NASA Exoplanet Archive\footnote{\url{https://exoplanetarchive.ipac.caltech.edu/}} \citep{akeson2013nasa}, which compiles data from missions such as $Kepler/K2$ \citep[]{borucki2010kepler}, the Transiting Exoplanet Survey Satellite \citep[$TESS$;][]{ricker2015transiting}, and ground-based instruments like $HARPS$ \citep[]{2011arXiv1109.2497M} and $ESPRESSO$ \citep[]{pepe2021espresso}. Cross-matching this table with our selected main-sequence hosts yields 1,471 confirmed planets orbiting 1,255 stars.  

We computed stellar luminosities by considering $Gaia$ data and following the formulation presented in \citet{pijpers2003selection}
\begin{equation}
\begin{split}
 \text{log}_{10} (L/L_{\odot}) &= 4.0 + 0.4 M_{\text{bol},\odot} - 2 \times \log_{10} \pi(\text{mas}) \\
&\quad - 0.4 \left( G - A_{\text{G}} + BC \right),
\end{split}
\label{lum_eqn}
\end{equation}
where $\pi$ and $G$ represent $Gaia$ parallax and apparent $G$-band magnitude, respectively \citep{2023Colla}. Our targets have $Gaia$ DR3 $G$-band magnitudes between 6 and 21, except for 55 Cancri (55 Cnc), which has $G=5.73$. We corrected the parallaxes for the zero-point offset using the Python code\footnote{https://gitlab.com/icc-ub/public/gaiadr3 zeropoint} provided by \citet{lindegren2021gaia}, which models the offset as a function of magnitude, colour, and position. For 55 Cnc, a value of -0.029 was adopted, as recommended by \citet{lindegren2018gaia}. The bolometric corrections, $BC$, were determined using the coefficients provided by \citet{andrae2018gaia}. We adopted a solar bolometric magnitude of $M_{\text{bol},\odot} = 4.74$, following the IAU recommendation by \citet{2015arXiv151006262M}. To account for interstellar absorption, we calculated the $G$-band extinction using
\begin{equation}
    A_{\text{G}} = R\times E(B - V)~, 
\end{equation}
 where $E(B - V)$ is the colour excess, obtained using the dust maps Python package\footnote{\url{https://dustmaps.readthedocs.io/en/latest/}} by \citep{2018JOSS....3..695M}, which queries three-dimensional reddening maps. An extinction coefficient of $R = 2.74$ for the $G$-band \citep[][]{casagrande2018use} was adopted. The uncertainties in the derived luminosities were determined through standard error propagation incorporating the uncertainties in parallax and the $G$-band magnitude. We note that stars with Renormalised Unit Weight Error \citep[RUWE:][]{lindegren2018gaia} values above $\sim 1.4$ are identified as having unreliable astrometric solutions, usually due to flux contamination from binarity, and thus we exclude them.

 The trained MAISTEP model was then applied to characterise 1,221 stars (hosting 1,405 planets), yielding estimates of their radii, masses, and ages. In MAISTEP, input uncertainties are propagated by random sampling, assuming Gaussian errors in  $T_{\text{eff}}$, [Fe/H], and $L$. We use the sample median as the central estimate and the 68\% confidence interval to define parameter uncertainty. The inferred radii, masses, and ages achieve median fractional uncertainties of 2\%, 2\%, and 27\%, respectively.  
 The updated stellar radius ($R_{\star}$), combined with star-to-planet radius ratio ($\rho$) from the NASA Exoplanet Archive, were used to revise the planet radius ($R_{\text{P}}$), following the expression:
 \begin{equation}
      \rho = \frac{R_{\text{p}}}{R_\star}. 
 \end{equation}
We calculated the uncertainty in $R_{\text{P}}$ using the standard error propagation approach. The planet-to-star radius ratio is derived using transit models that incorporate limb darkening coefficients, which themselves depend on stellar parameters such as effective temperature, surface gravity, and metallicity \citep[e.g.,][]{neilson2022limb}. While a fully consistent re-derivation would likely alter these values, we assume the impact is negligible from a statistical perspective, given the modest variation in stellar parameters across our sample.

Following \citet{fulton2017california}, we restrict the sample to systems with orbital periods $<$ 100 days, where transit surveys achieve high completeness for small planets ($\sim 4$ R$_\oplus$) and reliable transit detection. As noted by \citet{petigura2020two}, relative uncertainties in planet-to-star radius ratios exceeding $\sim10$\% -- especially for high-impact-parameter transits ($b \gtrsim0.8$), where degeneracies with limb darkening are strongest -- can blur the location of the radius valley. In our analysis, we exclude planets with relative uncertainties in the planet–star radius ratio greater than 10\%. Further, to focus on the range where the valley is located, the super-Earths and sub-Neptunes region, we limit the sample to planets with radii between 1 and 4 R$_\oplus$. This results in a sample of 893 confirmed planets hosted by 779 stars. The median uncertainties for stellar hosts in our sample are 60 K in $T_{\text{eff}}$, 0.06 dex in [Fe/H], and 0.02 L$_\odot$ in luminosity. This set represents the complete sample of host stars and transiting planets used in our analysis. Owing to the high precision of the stellar input parameters and the planet–star radius ratios, any heterogeneity within the sample is expected to have minimal impact on our inferred results.

\begin{figure}[t]
        \centering
        \includegraphics[scale=0.4]{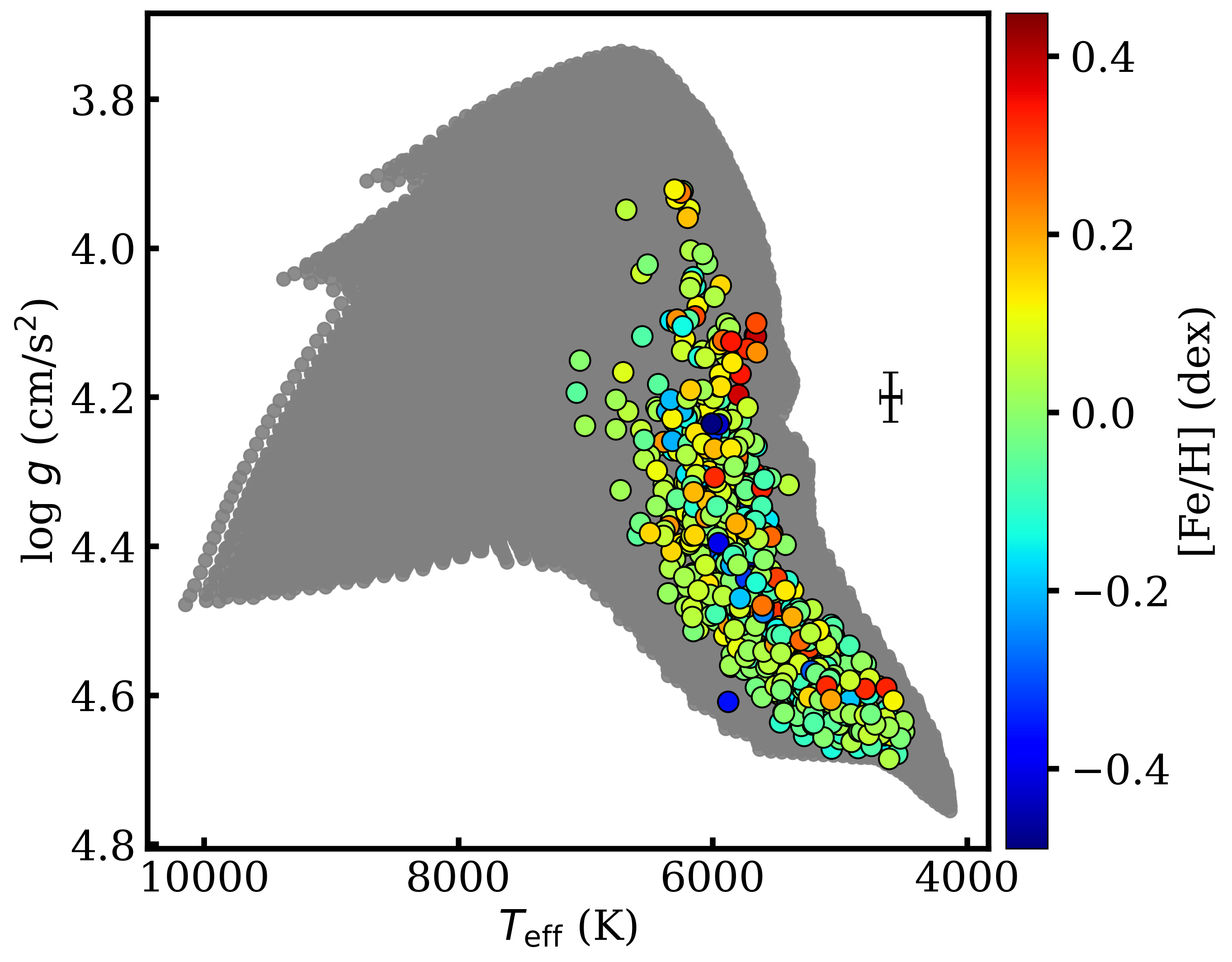}
            \caption{Kiel diagram showing 779 selected planet host stars overlaid on the main-sequence model grid (gray points). The marker shows median uncertainties of 0.05 dex in log $g$ and 60 K in $T_{\text{eff}}$.}
            \label{kiel}
\end{figure}

Among the selected sample, 820 planets orbit 724 stars observed by $Kepler/K2$, 64 planets orbit 53 hosts observed by $TESS$, and 9 planets around 2 stars were discovered using multiple observation facilities. The 779 host star sample is shown in the Kiel diagram in Fig.~\ref{kiel}. These stars lie within the surface gravity--effective temperature plane spanned by the MAISTEP training models and have effective temperatures between  4400 -- 7100 K, corresponding to FGK spectral types. A comparison with datasets from  \citet{fulton2018california}, \citet{2023arXiv230111338B}, and \citet{silva2015ages} is presented in Appendix~\ref{appendix}. Overall, the inferences show good agreement, with median offsets (and scatter) of  $\sim$4\% (5\%) in radius and 2\% (6\%) in mass. For stellar age, we obtain a median offset of $\sim$7\% but a much larger scatter of 93\%. The differences -- particularly in radius -- arise in part from the adopted zero-point parallax corrections. On the other hand, the differences in stellar age estimates may be attributed to differences in the  physics used in the stellar evolutionary models.

\section{The radius valley}
\label{gap}
\begin{figure}[t]
        \centering
        \includegraphics[scale=0.45]{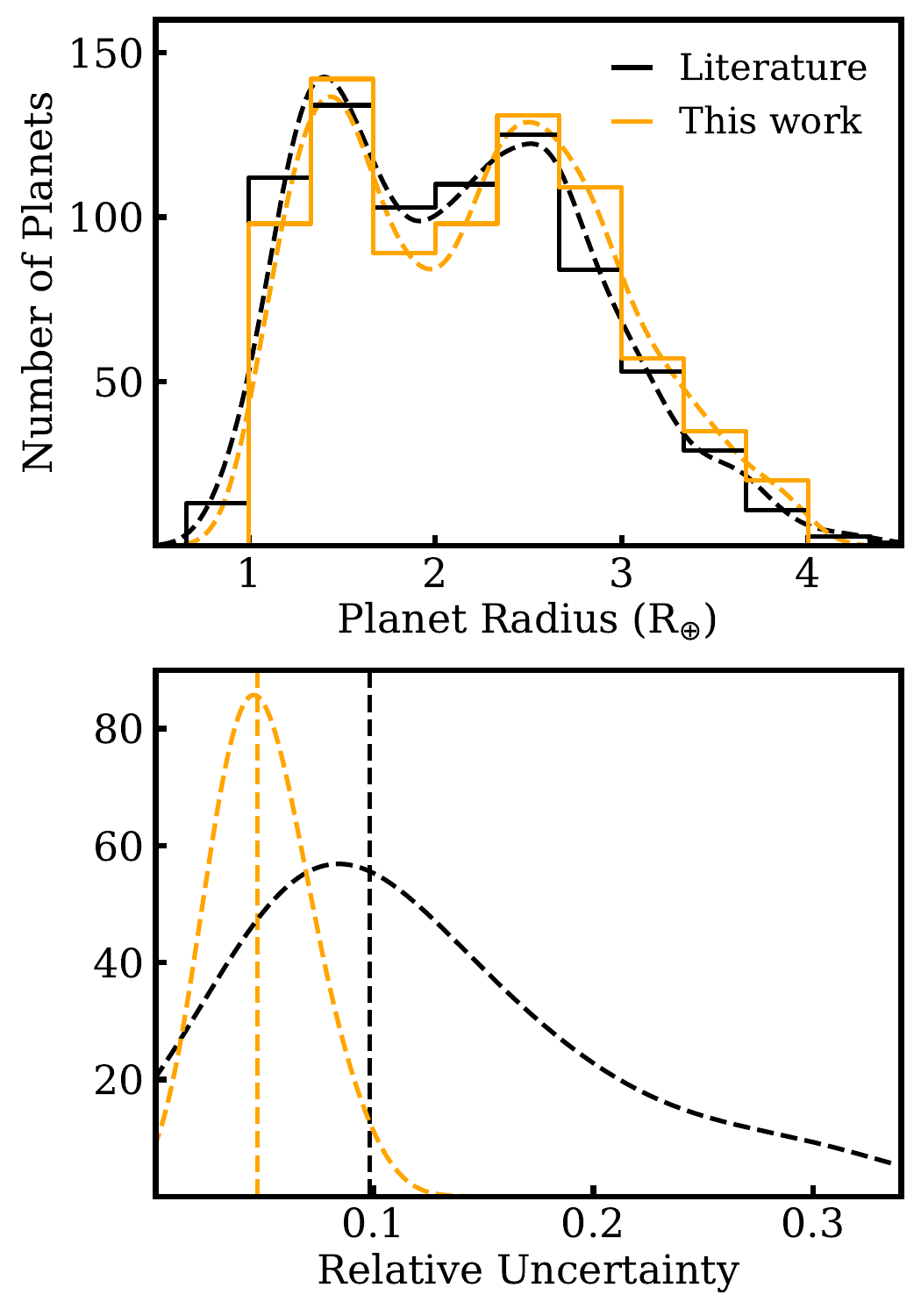}
            \caption{Top panel: Histograms of planet radii from the literature (black) and our revised values (orange), using 0.33 R$_{\oplus}$ bin width. Dashed curves represent kernel density estimates (KDEs) computed with 0.33 R$_{\oplus}$ bandwidth. Lower panel: KDE distributions of radius uncertainties, with vertical dashed lines indicating the medians: 0.05 (this work) and 0.10 (literature values from the NASA Exoplanet Archive).}
            \label{pl_radius_distributions}
\end{figure}
Figure.~\ref{pl_radius_distributions} (top panel) shows the planet radius distribution based on both our revised estimates and literature values from the NASA Exoplanet Archive. The revised distribution (orange line) displays a clear bimodal structure, with a valley around 2 R$_\oplus$ and peaks near $\sim$1.4 R$_\oplus$ and $\sim$2.5 R$_\oplus$. This bimodality is evident in both the histogram and the kernel density estimate, KDE (estimated by placing a Gaussian kernel/function at each data point and summing them to produce a continuous distribution) calculated using a bandwidth of 0.33 R$_\oplus$. For comparison, the distribution based on the literature values compiled in the NASA Exoplanet Archive is shown in black.

Our updated planetary radius measurements suggest that the radius valley is only partially populated. This aligns with the interpretation of \citet{fulton2018california}, who argued that the valley is not entirely devoid of planets. \citet{van2018asteroseismic}, using a sample of 117 planets orbiting stars with precisely measured radii from asteroseismology, found the valley to be wider and more depleted. They proposed that the valley may be intrinsically empty and that the apparent infill in other studies could stem from larger radius uncertainties. Since our typical planetary radius precision is comparable due to the high level of precision of the MAISTEP stellar inputs (relative median uncertainty of $\sim 5$\%, see Fig.~\ref{pl_radius_distributions}), their conclusion may have been influenced by the limited sample size.
 
The top panel of Fig.~\ref{pl_radius_distributions} also shows that our revised distribution yields a well-defined and relatively deeper valley than that based on literature values. The depth of the radius valley has been shown in previous studies to correlate with data quality, specifically, the reliability of planetary radii, which in turn depends critically on the host star radius \citep[e.g.,][]{fulton2017california,fulton2018california,ho2023deep,2025A&A...699A.100D}. Our method, described in Sect.~\ref{method}, provides stellar radius estimates of median precision of 2\%, leading to planetary radii with a median relative uncertainty of 5\%, a significant improvement over the 10\% uncertainty derived from the values in the NASA Exoplanet Archive, as shown in the lower panel of Fig.~\ref{pl_radius_distributions}.

To quantify the depth of the valley, \citet{fulton2017california} introduced the metric $V_{\text{A}}$, defined as:
\begin{equation}
   V_{A} = \frac{N_{\text{valley}}}{{(N_{\text{first}} \times N_{\text{second}})^{0.5}}}, 
\end{equation}
where $N_{\text{valley}}$, $N_{\text{first}}$, and $N_{\text{second}}$ represent the estimated number of planets in the valley, first peak, and second peak, respectively. In this analysis, these values were determined based on the radius ranges of $1.67- 2.32$ R$_{\oplus}$, $1.35 - 1.67$ R$_{\oplus}$, and $2.32 - 2.8$ R$_{\oplus}$, respectively. A value of $V_{\text{A}} = 1$ corresponds to a log-uniform distribution, while lower values indicate a more pronounced valley. Using our updated radii, we find $V_{\text{A}} = 0.686$, compared to $V_{\text{A}} = 0.824$ based on values in the NASA Exoplanet Archive, confirming that our revised valley is deeper. 

 In the next sections, we examine how the location of the radius valley varies with orbital and stellar parameters.

\subsection{On the orbital period dependence}
\label{pl_rad_orbper}
\begin{figure}[t]
        \centering
        \includegraphics[scale=0.4]{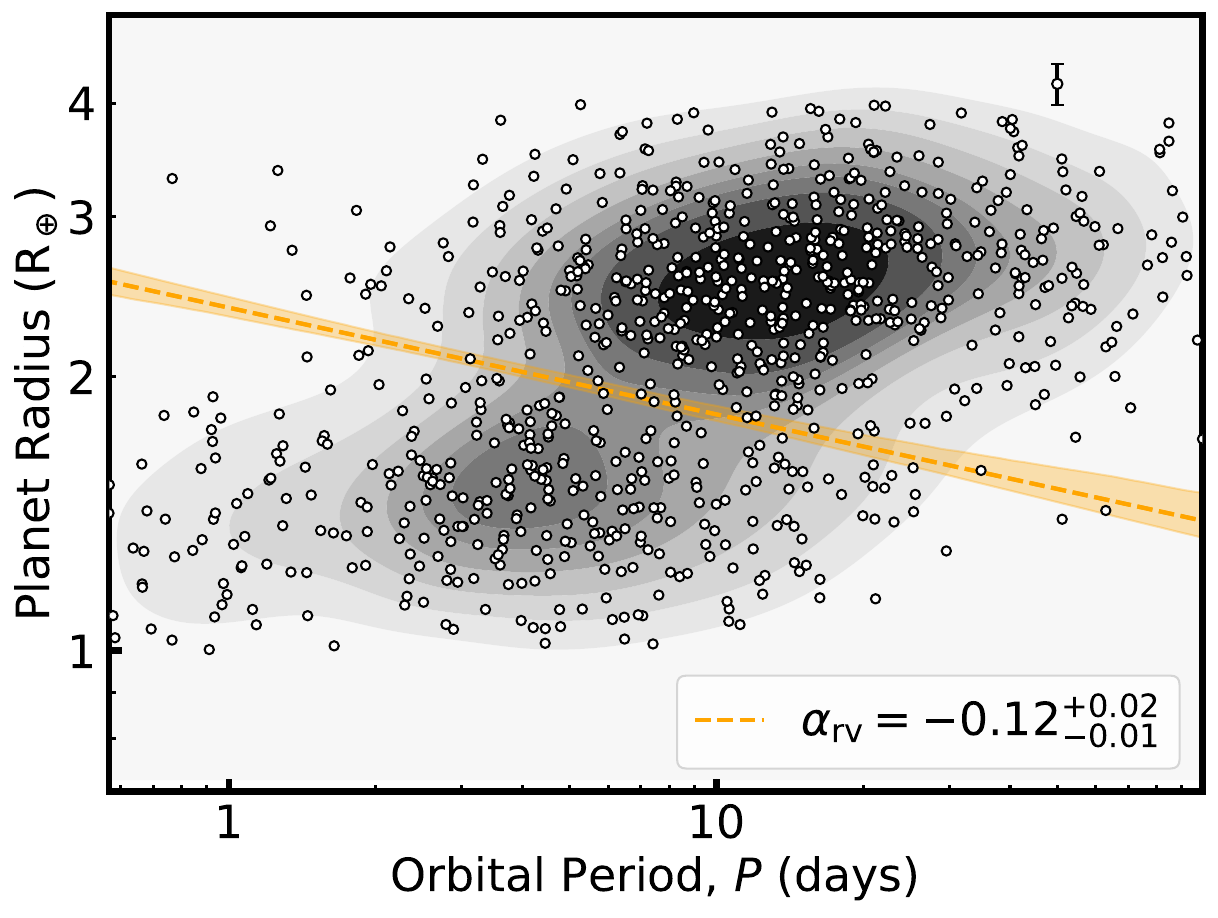}
            \caption{Planet radius against orbital period. Small circles represent individual planet detections. Contours show the two-dimensional KDE of the planet population. The orange line is the best-fit trend, with a shaded band indicating the 68\% confidence interval from bootstrap sampling. The error bar indicate the median uncertainty in planet radius.}
            \label{pl_rade_orbper}
\end{figure}

Figure~\ref{pl_rade_orbper} shows the planets in our sample on a period–radius diagram, where the radius valley appears as a gap in the distribution, sloping downward towards longer orbital periods. Given that the slope of the valley carries information about the physical mechanisms responsible for shaping it, we quantified this feature using the \texttt{gapfit}\footnote{https://github.com/parkus/gapﬁt} tool  \citep{loyd2020current}. In brief, the approach involves testing various straight-line models that describe how the valley depends on an independent parameter, such as orbital period. For each model, the predicted valley radius is computed individually for each planet based on its orbital period, and this value is subtracted from the planet's observed (inferred) radius. This centers the valley at the zero line on the transformed radius axis, serving as the origin for measuring radius deviations. KDE is then applied to evaluate the number density of planets near the gap. The best-fit valley model is the one that minimizes the planet density at zero in the transformed space. Uncertainties in the fitted gap parameters are estimated by bootstrap resampling of the observed data. Each iteration randomly samples these points with replacement, fits the gap parameters to the resampled set, and builds a distribution of fits to quantify the parameter uncertainties.

We model the valley as a linear trend in the log-log space:

\begin{equation}
    \text{log}_{10} (R_{\text{p}}) = \alpha_{\text{rv}}\times(\text{log}_{10}(P) - \text{log}_{10}(P,0)) + \text{log}_{10} (R_{\text{p,0}}),
\end{equation}
where $\text{log}_{10}(R_{\text{p}})$ is the planet radius, $\text{log}_{10}(P)$ is the orbital period, and the valley location is parameterised by the slope $\alpha_{\text{rv}}$ and the the intercept, $\text{log}_{10}(R_{\text{p,0}})$. $\text{log}_{10}(P,0)$ is a reference fixed orbital period, chosen near the center of the data to minimize the correlation between slope and intercept in the fit, following \citet{loyd2020current}.

The planet sample was further restricted to only those with radii between $1.4-2.6$ R$_{\oplus}$, to focus the fit primarily on the distribution within the valley. The initialization parameters were: $\text{log}_{10}(P,0) = 1$, representing the reference period in the gap line equation; an initial guess for the planet size, $\text{log}_{10}R_{\text{p,0}} = 0.3$; an initial slope guess of $\alpha_{\text{rv}} = -0.15$; and a width $\sigma = 0.15$ for the KDE used in the fitting. We allowed the valley line to vary within a range of $\pm 0.5$ from the initial $\text{log}R_{\text{p,0}}$, and the analysis included $n =1000$ bootstrap simulations with replacement using the Nelder-Mead simplex optimization method \citep[][]{nelder1965simplex}. The sensitivity of the fits to the choice of these parameters is discussed in detail by \citet{loyd2020current}. From the resulting fits, we extracted the median and the 16th and 84th percentiles to define the 68\% confidence interval. The best-fit line and its uncertainty envelope, shown as an orange-shaded region below and above the line in the plot, indicate a trend in the radius valley with a slope $\alpha_{\text{rv}} = -0.12^{+0.02}_{-0.01}$. 

    This slope aligns with previous observational studies \citep[e.g.,][]{van2018asteroseismic,fulton2018california,martinez2019spectroscopic, petigura2022california,ho2023deep} and has been explained by evolution models based on photoevaporation or core-powered mass loss. In photoevaporation, the mass-loss timescale for planets with small envelopes decreases during evaporation, leading to complete stripping, while it increases for larger envelopes, peaking when the planet radius roughly becomes twice the core size. As a result, some planets retain substantial envelopes that inflate their radii, while others are stripped down to bare cores. Higher core-mass planets resist evaporation, raising the transition radius at short periods. At longer periods, even low-mass planets can retain envelopes, shifting the radius valley and rocky planet sizes downward. The predicted slopes of the valley by photoevaporation are in the range between -0.1 $\leq \alpha_{\text{rv}} \leq $-0.25, depending on the evaporation  efficiency and the choice of model, with numerical simulations typically producing shallower trends than analytical estimates \citep[][]{owen2013kepler,owen2017evaporation}. 
    
    Under core-powered mass loss, the planet’s equilibrium temperature, $T_{\text{equ}}$, determines its atmospheric cooling timescale, and the mass-loss timescale has an exponential dependence on both orbital period and planet size. As planets lose mass, they can also contract, which increases the mass-loss timescale. At some point, the two timescales match, a condition which determines whether a planet ends up above or below the valley. Since $P \propto T_{\text{equ}}^{-3} $, planets at longer orbital periods (lower $T_{\text{equ}}$) cool more slowly, allowing even small cores to retain their envelopes. Using an analytical model within this framework, \citet{gupta2019sculpting} derived a slope of -0.11.
    
    Scenarios involving the late formation of rocky planets in gas-poor environments predict that, under the assumed radial power-law of slope -1.5 for the planetesimal disk surface density, the amount of solid material increases with distance from the star, leading to a transition radius trend with a positive sign \citep[e.g.,][]{lee2014make,lee2016breeding,lopez2018formation,lee2019episodic}. \citet{venturini2024fading} predicted a slope of -$0.16^{+0.01}_{-0.01}$ using a hybrid model of planet formation and photoevaporation.

Two other regions in Fig.~\ref{pl_rade_orbper} exhibit a notable scarcity of planets. First, there is a deficit of sub-Neptune-sized planets ($\gtrsim 2$ R$_\oplus$) at very short orbital periods ($\lesssim 2$ days). Given that planets of this size would be readily detectable in this parameter space, the observed lack is thought to be physical rather than due to observational bias. This region of low sub-Neptune occurrence is referred to as the sub-Neptune desert/savanna \citep[e.g.,][]{szabo2011short,beauge2012emerging,lundkvist2016hot,west2019ngts,bourrier2023dream,castro2024mapping}. Secondly, the apparent scarcity of super-Earth-sized planets at longer orbital periods is primarily attributed to reduced detection efficiency and survey completeness in this regime \citep[see e.g.,][]{petigura2022california}. At these separations, small planets produce fewer and shallower transits, making them more difficult to detect in transit surveys such as $Kepler$. Although the radius valley is commonly analysed using orbital period, we also explore its dependence on incident stellar flux.

\subsection{On the incident flux dependence}
\begin{figure}[t]
        \centering
        \includegraphics[scale=0.4]{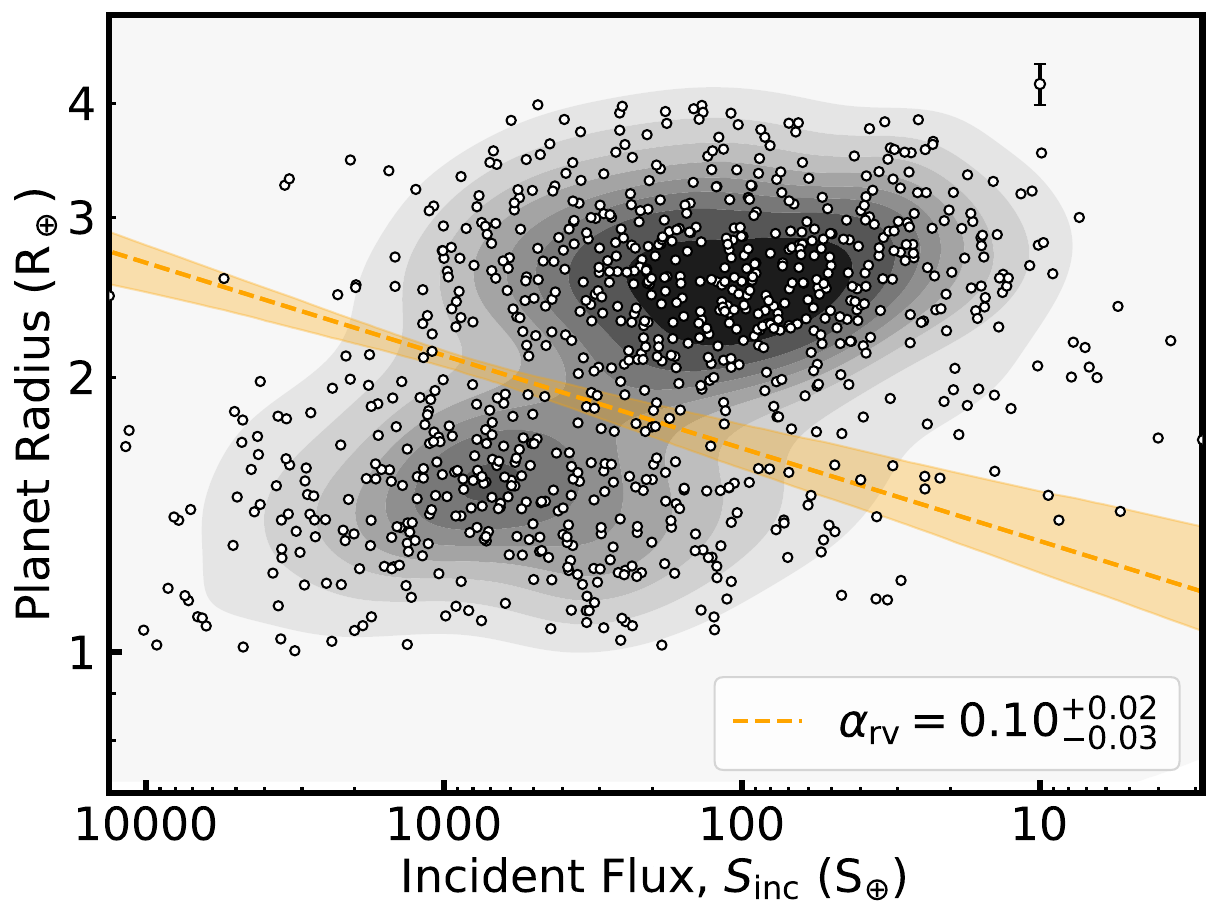}
            \caption{Same as Fig.~\ref{pl_rade_orbper}, except for planet size against incident flux. See text for details}
            \label{pl_rade_inc_flux}
\end{figure}
We compute the stellar flux incident on a planet by assuming flux conservation under isotropic radiation, as described by the inverse-square law:
\begin{equation}
    S_{\text{inc}} = \frac{L_{\star}}{4\pi a^{2}},
    \label{inc_flux}
\end{equation}
where $L_{\star}$ is the luminosity of the host star, calculated using Eq.~\ref{lum_eqn}, and $a$ is the semimajor axis of the planet's orbit. Assuming circular orbits and applying Kepler’s law, $a$ is determined as follows:

\begin{equation}
    a^{3} = \frac{GM_{\star}P^{2}}{4\pi},
    \label{semimajor}
\end{equation}
where $P$ denotes the orbital period provided by the NASA Exoplanet Archive. We assumed circular orbits because only 119 planets have measured orbital eccentricities.

In Fig.~\ref{pl_rade_inc_flux}, we present the distribution of planets on an incident flux-radius plane. Using the same fitting procedure described in Sect.~\ref{pl_rad_orbper}, we fitted the radius valley, modifying only the initial guess for the slope and reference point in incident flux to account for the observed trend and the change in the x-axis parameter. 
Unlike in the case with orbital period, the slope here is positive, with a value $\alpha_{\text{rv}} = 0.10^{+0.02}_{-0.03}$ (note that the incident flux in the horizontal axis is reverted to keep a similar distribution of the points. This explains the change of sign). This implies that the incident flux scales with the orbital period according to $S_{\text{inc}} \propto P^{-4/3}$ for a given host star mass (see Eq.~\ref{inc_flux} \&~\ref{semimajor}). Such a trend has also been reported by \citet[][]{fulton2018california}, \citet{martinez2019spectroscopic}, and \citet{ho2023deep}. Figure.~\ref{pl_rade_inc_flux} also shows a scarcity of sub-Neptunes at high incident flux and super-Earths at low flux, as previously discussed in the context of orbital period (Sect.~\ref{pl_rad_orbper}). Owing to the luminosity–mass relation for main-sequence stars, we also investigate the dependence of the valley on host star mass.

\subsection{On the host-star mass dependence}
\label{mass_dependence}
 
\begin{figure}[t]
        \centering
        \includegraphics[scale=0.4]{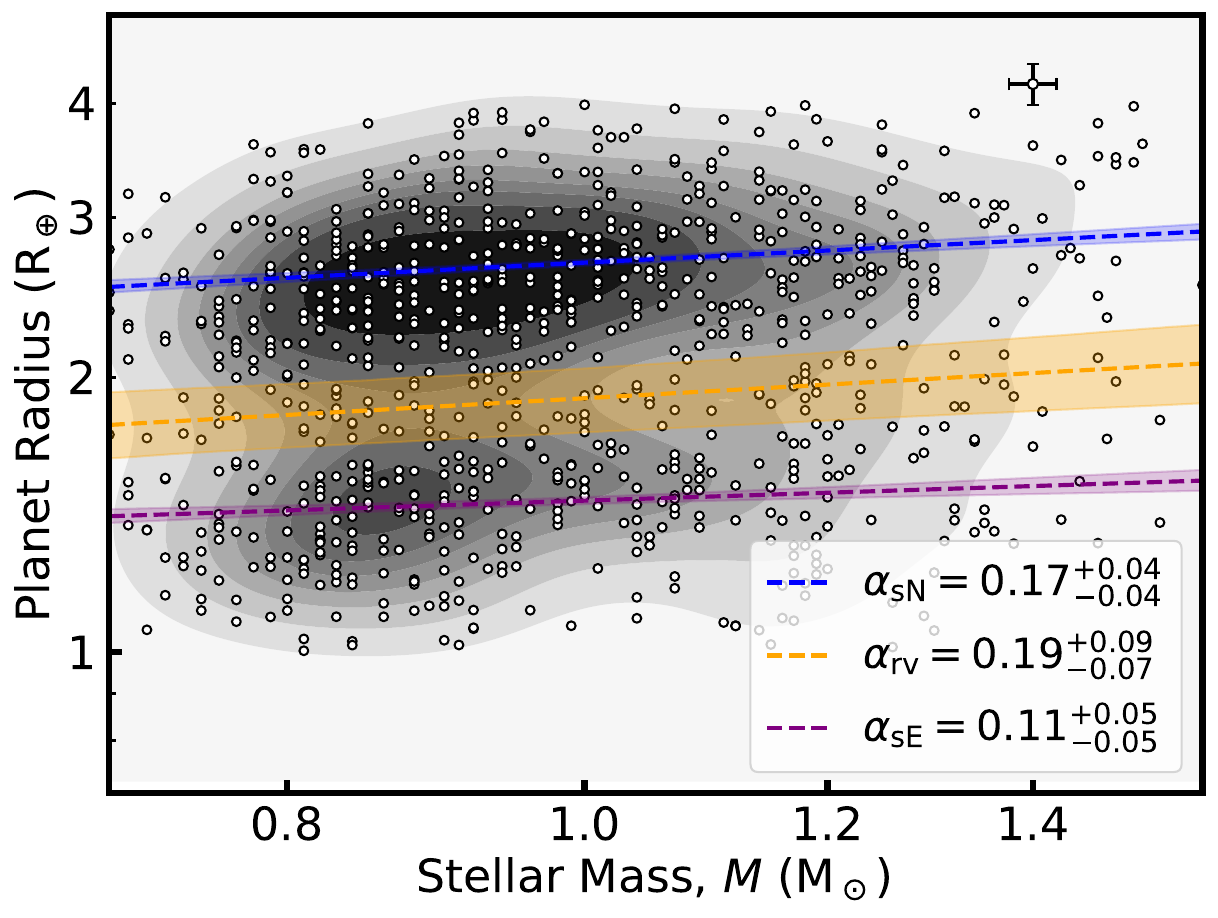}
            \caption{Same as Fig.~\ref{pl_rade_orbper}, but for planet size as a function of host star mass. See text for details.}
            \label{pl_rade_star_mass}
\end{figure}

Figure~\ref{pl_rade_star_mass} shows that the valley appears at larger radius with increasing stellar mass, with a slope of $\alpha_{\text{rv}} = 0.19^{+0.09}_{-0.07}$. This finding agrees with that of \citet{berger2020gaiab}, who reported a slope of $0.26^{+0.21}_{-0.16}$ (our measurement has small errors), and is consistent within 1$\sigma$ that of \citet{ho2023deep}, $\alpha_{\text{rv}} = 0.23^{+0.09}_{-0.08}$. Our inferred slope also aligns with predictions from photoevaporation and core-powered mass loss evolution models. In a photoevaporation model, \citet{owen2017evaporation} and \citet{rogers2021photoevaporation} predicted slopes of 0.32 and 0.29, respectively. \citet{wu2019mass} also showed that reproducing the observed stellar-mass dependence of the radius valley within a photoevaporation framework requires assuming a linear scaling between planet and host mass, proposing that the peak of the planet mass distribution scales with stellar mass as $M_{\text{P}} \propto M_{\star}^{\beta}$, with $\beta$ in the range [0.95–1.40]. This implies a radius scaling of $R_{\text{P}} \propto M_{\star}^{\alpha}$, where $\alpha$ falls between 0.23 and 0.35, encompassing our inferred slope within 1$\sigma$ ($0.19^{+0.09}_{-0.07}$). Photoevaporation, driven by the total high-energy XUV radiation from stellar activity, scales  as $\propto M_{\star}^{-3}$ \citep[][]{jackson2012coronal}. This would prevent close-in planets around low-mass stars from retaining their envelopes. Therefore, sub-Neptunes around these stars are mostly found on wide orbits, while closer-in planets are stripped down to rocky cores.

\citet{gupta2020signatures} in a core-powered mass-loss model, predicts a slope of $\sim 0.35$ based on numerical simulations, and a slightly shallower slope of $\sim 0.33$ from analytical estimates, assuming a fixed period distribution and employing the bolometric luminosity-mass relation. The similarity in the predicted slopes of both models suggests that our measured slope of $0.19^{+0.09}_{-0.07}$ does not favour one model over the other. In a hybrid model of formation and photoevaporation by \citet{venturini2024fading}, which considers an extended stellar mass range of $0.1-1.5$ M$_\odot$, a shallow slope of $0.14^{+0.02}_{-0.01}$ was predicted. 

Another feature, evident from the KDE contours in Fig.~\ref{pl_rade_star_mass}, is the systematic shift in the peaks of both planetary populations on either side of the valley towards larger radii with increasing stellar mass. We quantify this trend by classifying the sample into super-Earths and sub-Neptunes using a radius split at 1.98 R$_\oplus$, which is the derived centroid of the valley in one-dimension determined using a Gaussian model mixture (GMM) grouping (see also Fig.~\ref{pl_radius_distributions}). 
We performed bootstrap linear regression within each group to derive power-law relations between planet radius and stellar mass, as shown in blue and purple in Fig.~\ref{pl_rade_star_mass}. We find slopes of $\alpha_{\text{sN}} = 0.17^{+0.04}_{-0.04}$ for sub-Neptunes and $\alpha_{\text{sE}} = 0.11^{+0.05}_{-0.05}$ for super-Earths, respectively. The steeper slope for sub-Neptunes indicates a stronger dependence of planet size on stellar mass for this sample compared to super-Earths. The difference in slopes may suggest that higher-mass stars slightly inflate the radii of their planets, leading to a steeper slope in the sub-Neptune population, but not in the bare, rocky super-Earths. Generally, these results highlight that planets orbiting stars in our study mass range (0.7–1.6 M$_\odot$) tend to be systematically larger around more massive stars, with the radius valley location changing to larger planets as stellar mass increases. This trend likely reflects the dependence of core size growth on the amount of solid material in the protoplanetary disk, which scales with stellar mass \citep[e.g.,][]{andrews2013mass,pascucci2016steeper,ansdell2017alma}. Consequently, this scaling influences not only planets within the valley, but also those on either side encompassing rocky super-Earths and gaseous sub-Neptunes \citep[e.g.,][]{wu2019mass}. 

A similar trend is visible in the figures of \citet[][Fig.~8]{fulton2018california} and \citet[][Fig.~5]{berger2020gaiab}, both showing a positive $R_{\text{P}}$–$M_{\star}$ correlation for sub-Neptunes, super-Earths, and the valley population across stellar mass ranges of roughly $0.8-1.3$ M$_{\odot}$ and $0.5-1.5$ M$_{\odot}$, respectively, although the trend was not quantified statistically. In a  core-powered mass-loss scenario, \citet{gupta2020signatures} attribute the observed trend in both the radius valley and the two populations to the assumption that the orbital period distribution is assumed to be independent of stellar mass \citep[motivated by observation data in][]{fulton2018california}. As a result, planets at the same orbital period will have higher $T_{\text{equ}}$ around more massive (i.e., more luminous) stars. For a constant orbital period, $T_{\text{equ}} \propto M_{\star}^{0.25(\alpha-2/3)}$, with $\alpha$ being the power-law index in the stellar mass-luminosity relation. The higher $T_{\text{equ}}$ enhances the atmospheric escape, pushing the radius valley to larger sizes and shifting both the sub-Neptunes and super-Earths to higher $T_{\text{equ}}$ regions.

\citet{petigura2022california} applied a 1.7 R$_\oplus$ radius cut to planets orbiting stars in the $0.5-1.4$ M$_\odot$ range and reported a stellar mass dependence for sub-Neptunes, with a slope of $0.25^{+0.03}_{-0.03}$, consistent within 1$\sigma$ with our estimate of $0.19^{+0.05}_{-0.05}$. They did not find a comparable trend for super-Earths, reporting a much shallower slope of $ 0.02 ^{+0.03}_{-0.03}$, which is notably lower than the values we derive. This slope is also lower compared the apparent trend in \citet{fulton2017california}, \citet{berger2020gaiab} and the inference of $0.23-0.35$ from \citet{wu2019mass}.  \citet{petigura2022california} suggested that more massive protoplanetary disks around higher-mass stars allow sub-Neptune cores to accrete thicker gaseous envelopes, resulting in larger observed radii. These cores grow roughly linearly with stellar mass, following $M_{\text{c}} \approx 10 M_{\oplus}(M_{\star}/M_{\odot})$, and can exceed the threshold mass above which atmospheric stripping becomes inefficient. As a result, many sub-Neptune planets retain substantial envelopes despite ongoing loss processes. In contrast, super-Earths are likely stripped cores that fall below this threshold. Because the radii of rocky planets increase only weakly with core mass, their sizes remain nearly constant across a range of host masses. Therefore, the apparent increase in planet size with stellar mass for super-Earths in Fig.~\ref{pl_rade_star_mass} may result from the slightly higher masses for a subset of stars in this work compared to values reported in the literature (see Fig.~\ref{MAISTEP_SPInS} in the Appendix).
 
The main-sequence lifetime of a star depends strongly on its mass, approximately following $\tau_{\text{ms}} \sim 10~\text{Gyr}~(M_{\star}/\text{M}_{\odot})^{-2.5}$ \citep[][]{kippenhahn1990stellar,kippenhahn2012}. This mass–age correlation implies that the radius valley may also depend on the age of the star-planet system. In the following section, we use stellar age estimates from MAISTEP to investigate the temporal evolution of features in the small-sized exoplanet population.

\subsection{On the host-star age dependence}
\label{age_dependence}

The age of a planetary system is typically inferred from that of its host star, as planets form during the early stages of stellar evolution \citep[][]{meyer2008circumstellar}. However, stellar age is not directly observable and remains one of the most difficult stellar parameters to determine, especially for field stars, with no single method performing reliably across all stellar types and evolutionary stages. Asteroseismology provides highly precise stellar age estimates, but the number of main-sequence stars with seismic detections and confirmed exoplanets remains $<$ 100 \citep[][]{silva2015ages,campante2015ancient,lundkvist2016hot,lund2017standing,campante2019tess}. This is expected to change with the upcoming ESA's $PLATO$ mission \citep{rauer2014plato}. 

\begin{figure}[t]
        \centering
        \includegraphics[scale=0.4]{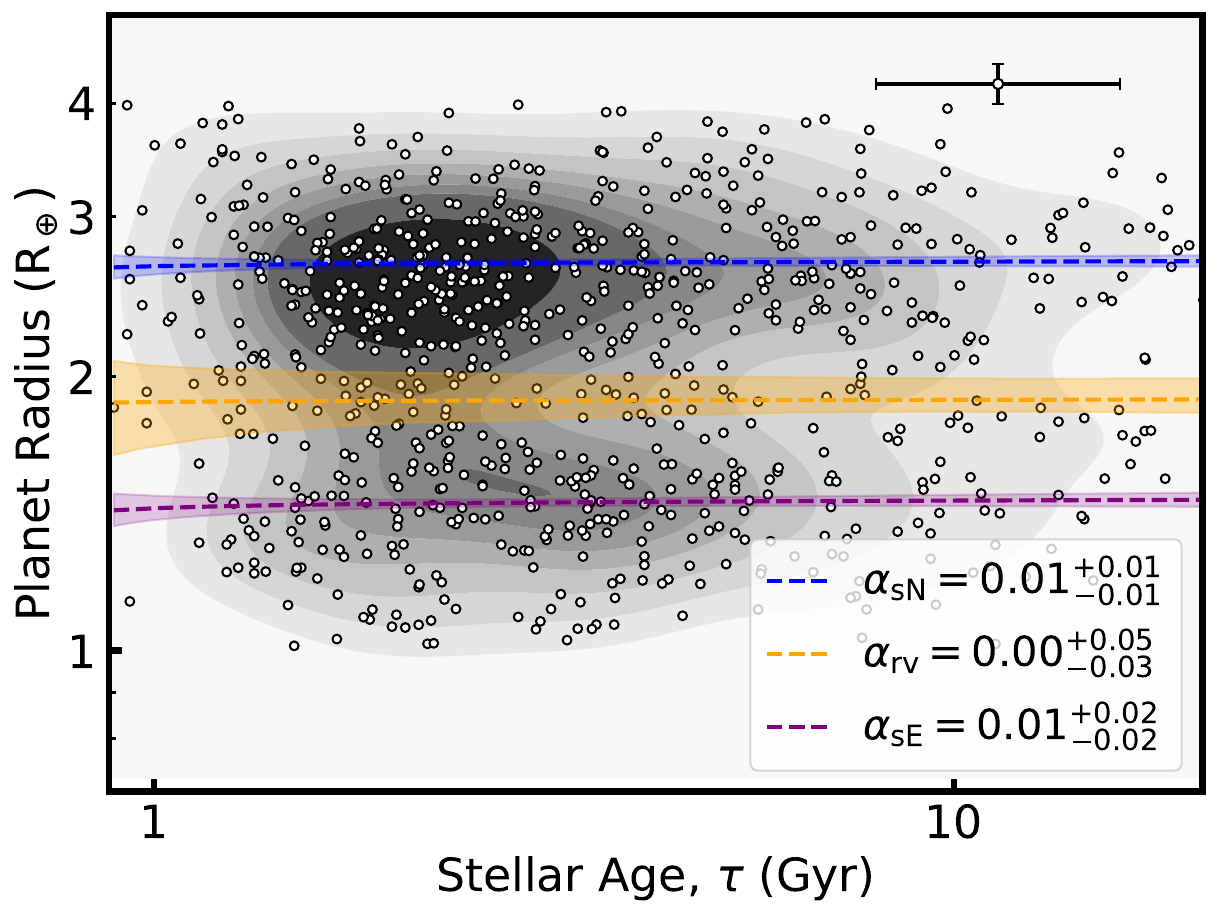}
            \caption{Same as Fig.~\ref{pl_rade_orbper}, but for planet size as a function of host star age. See text for details.}
            \label{pl_rade_star_age}
\end{figure}

In \citet{kamulali2025maistep}, we assessed stellar age estimates from MAISTEP, derived using only atmospheric constraints, by comparing them to asteroseismic ages for a sample of $Kepler$ legacy main-sequence stars. The MAISTEP stellar ages relative to the asteroseismic ages showed a bias of 7\%, a scatter of 23\%, and an average fractional uncertainty of 28\%. Using the same methodology, we estimated the ages of host stars in this study, achieving an average fractional uncertainty of 23\%. To minimize the impact of poorly constrained ages, we excluded 46 host stars with flat age distributions, resulting in a final sample of 823 planets orbiting 718 unique hosts. The removed stars typically have median ages near the grid midpoint  ($\sim 7$ Gyr) with broad uncertainties ($\sigma_{\tau} \gtrsim 3.5 $ Gyr). They are low-mass ($\lesssim 1.0$ M$_\odot$), and have $T_{\text{eff}} \lesssim 5600$ K. Given their small number, excluding these objects has a minor impact on observed mass and age trends.

Figure~\ref{pl_rade_star_age} presents the distribution of planet radii as a function of stellar age. The radius valley shows minimal variation with age, and a power-law fit to its location yields a slope of $\alpha = 0.00^{+0.05}_{-0.03}$, indicating no significant age dependence. Using isochrone-based ages, \citet[][]{ho2023deep} similarly reported a shallow slope of $\alpha = 0.033^{+0.017}_{-0.025}$.

The variation of average sizes of planet populations above and below the radius valley with age is also examined. The resulting fits, shown in blue and purple in Figure~\ref{pl_rade_star_age}, with slopes of $\alpha_{\text{sN}} = 0.01^{+0.01}_{-0.01}$ and $\alpha_{\text{sE}} = 0.01^{+0.02}_{-0.02}$ for sub-Neptunes and super-Earths, respectively, reveals no significant dependence.

These findings are consistent with those of \citet{petigura2022california} based on isochrone ages. In contrast, \citet{chen2022planets}, employing a kinematic method based on the average age-velocity dispersion relation for the LAMOST-Gaia-Kepler sample, reported a decline in average sub-Neptune size with increasing age, which is consistent with theoretical expectations. Both photoevaporation and core-powered mass loss models predict a decrease in sub-Neptune size over time with a slope of $\alpha = -0.1$, while super-Earth sizes are expected to remain largely constant. The discrepancy between these predictions and the findings based on our age estimates or isochrones may reflect limitations in stellar age determinations, underscoring the need for follow-up analysis using asteroseismic ages when they become available for a wider sample of host stars.

An alternative approach to probing the time evolution of close-in exoplanets is to examine the ratio of super-Earths to sub-Neptunes, which we denote as $N_{\text{E}}/N_{\text{N}}$. As sub-Neptunes are thought to lose their atmospheres over time and evolve into super-Earths, this ratio offers a population-level tracer of evolutionary processes. In this context, an increasing $N_{\text{E}}/N_{\text{N}}$ with system age may reflect the cumulative effects of atmospheric erosion over time, providing a complementary perspective to more direct observational signatures such as the radius valley or individual mass-radius measurements.

\begin{figure}[t]
        \centering
        \includegraphics[scale=0.43]{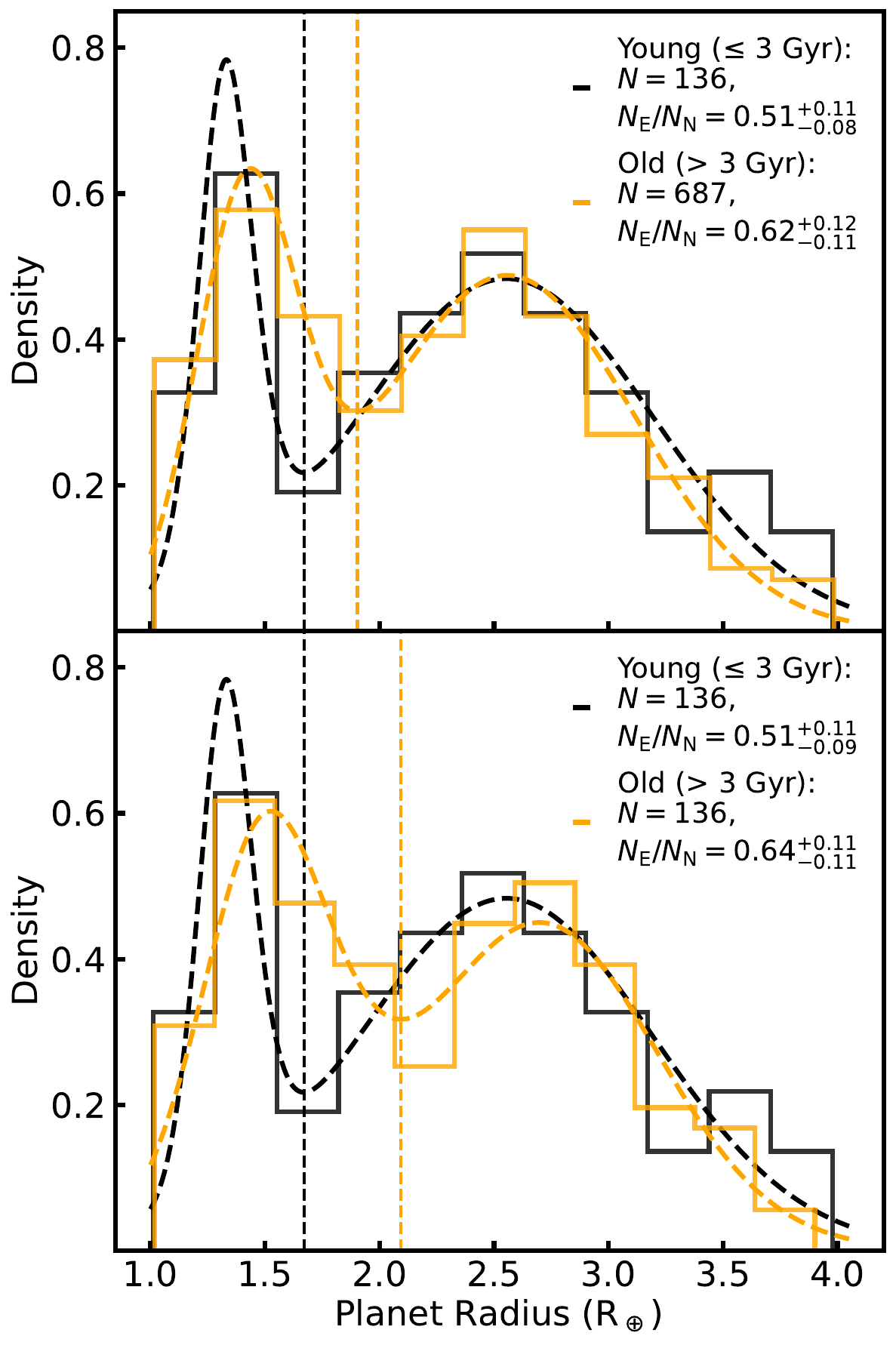}
            \caption{Radius distributions of young (black) and old (orange) planets. Top: full population. Bottom: property-matched sample in stellar metallicity, radius, and mass using optimal transport. Vertical dashed lines show the centroid of the inferred radius gap from GMM fits. Sample sizes, $N$ and ratios,  $N_{\text{E}}/N_{\text{N}}$ are indicated in panel titles.}
            \label{pl_rade_3Gyr_age_bins}
\end{figure}

We estimated the ratios by dividing the planetary sample into two age bins: systems younger than 3 Gyr and those older than 3 Gyr. We used 3 Gyr as the dividing point, as only 10 planets in our sample are hosted by stars with median ages below 1 Gyr. A Gaussian model mixture (GMM) was applied to identify the location of the radius valley in one dimension, after which we computed the number of planets below (super-Earths, $N_{\text{E}}$) and above (sub-Neptunes, $N_{\text{N}}$) the valley. The observational data were resampled 1,000 times with replacement to estimated the uncertainties in $N_{\text{E}}/N_{\text{N}}$. This approach yielded $N_{\text{E}}/N_{\text{N}} = 0.51^{+0.11}_{-0.08}$ for the younger systems and $N_{\text{E}}/N_{\text{N}} = 0.62 ^{+0.12}_{-0.11}$ for the older systems, as shown in the top panel of Fig.~\ref{pl_rade_3Gyr_age_bins}.  Additionally, the radius valley in the older population appears shallower and shifted to larger radii compared to the younger sample, based on the positions of the GMM-derived minima. The shift toward larger radii may reflect cumulative atmospheric loss in sub-Neptunes occurring over gigayear timescales. However, the large number of planets in the old sample (687 compared to 136 in the young population) may lead to a statistically biased result.

To examine potential sample size effects while accounting for age-related degeneracies, we selected a subset of older systems whose host metallicity, radius, and mass distributions closely resemble those of the younger population, with ages below 3 Gyr. We employ Optimal Transport \citep[OT:][]{flamary2021pot,flamary2024pot} in Python to identify a minimal-cost mapping between the joint distributions of stellar metallicity, radius, and mass for the two groups. Using the OT cost matrix, we compute a one-to-one assignment via the Hungarian algorithm \citep{kuhn1955hungarian}, yielding an old-star subset whose multivariate distribution closely reproduces that of the young sample. The Kolmogorov–Smirnov tests on the distributions of stellar metallicity, radius, and mass returned $p$-values of 0.95, 0.24, and 0.02, respectively, indicating that the young and old samples are statistically consistent in stellar metallicity and radius, with some differences in mass.
 As shown in the bottom panel of Fig.~\ref{pl_rade_3Gyr_age_bins}, the resulting ratio for the matched older sample is $N_{\text{E}}/N_{\text{N}} = 0.64^{+0.11}_{-0.11}$, which is consistent within uncertainties with the value of $0.62 ^{+0.12}_{-0.11}$ derived for the full older population. Therefore, at older ages, we find a higher prevalence of super-Earths relative to sub-Neptunes, suggesting that some sub-Neptunes evolve into super-Earths over time.
 
The change in the $N_{\text{E}}/N_{\text{N}}$ ratio was first reported by \citet{berger2020gaiab}, who used isochrone-based ages. They found that the ratio increased from $0.61 \pm 0.09$ for systems younger than 1 Gyr to $1.00 \pm 0.10$ for those older than 1 Gyr, focusing on hotter stars (5700–7900 K). A separate study by \citet{sandoval2021influence}, which included cooler GK-type stars ($T_{\mathrm{eff}} <$ 5800 K), explored age divisions at both 1 Gyr and 3 Gyr. Using the 3 Gyr cut, they reported $N_{\text{E}}/N_{\text{N}} = 0.78 \pm 0.17$ for younger systems and $1.00 \pm 0.04$ for older ones. Our results exhibit a similar age-dependent trend in $N_{\text{E}}/N_{\text{N}}$, albeit with somewhat lower absolute values. As  observed in \citet{david2021evolution} and \citet{ho2023deep}, we also find that the radius valley shifts to higher planetary radii and becomes shallower, indicating that the valley evolves over time. \citet{david2021evolution} suggested that the positive age trend arises from the scarcity of large super-Earths ($\sim 1.5-  1.8$  R$_\oplus$ ) orbiting young stars, a feature also evident in our Fig.~\ref{pl_rade_3Gyr_age_bins}.  

\subsection{Four-dimensional radius valley}
 A multidimensional radius valley has been proposed to better separate photoevaporation from core-powered mass loss. Here, we model the radius valley as a four-dimensional plane in the parameter space of planetary radius,  orbital period, stellar mass, and stellar age following the expression:
\begin{equation}
    \text{log}_{10}(R_{\text{P}}) = \alpha \text{log}_{10}(P) + \beta \text{log}_{10}(M_{\star}) + \gamma \text{log}_{10}(\tau) + \delta,
\end{equation}
 where $\alpha$, $\beta$, $\gamma$, are the coefficients (power-law indices) in orbital period, stellar mass, and age, respectively, and $\delta$ is the intercept on the planet radius axis.
 
 To achieve this, we use the logistic regression model implemented in Python via the $scikit-learn$ package. Unlike the gapfit method (see Sect.~\ref{pl_rad_orbper}), which is limited to two dimensions, logistic regression can be directly extended to higher dimensions. Logistic regression models the probability of a binary outcome from one or more predictor variables.  We first separate our sample into two groups, those above and below the valley, on the radius–orbital period–mass–age plane, using a two-component Gaussian Mixture Model. We identify the sub-Neptunes (sN) and super-Earths (sE) classes by comparing the mean log$_{10}(R_{\text{P}})$ of the two components; the cluster with the larger mean radius is taken as the sN population. To select an optimal regularisation parameter $C$ for logistic regression, we performed 5-fold cross-validation with GMM embedded in each fold to classify planets as sN or sE based on their radii distribution. Candidate $C$ values spanning 1 to 1000 were tested, and $C = 50$ maximised the mean classification accuracy across folds, and was used in the final model. Bootstrap with 1000 resamples of the 823 planets (with replacement) is performed to estimate the uncertainties in the coefficients. The posterior distributions of all coefficients are summarized using the median and the 16$^{\text{th}}$-- 84$^{\text{th}}$ percentile ranges.

 \begin{figure}[t]
        \centering
        \includegraphics[scale=0.38]{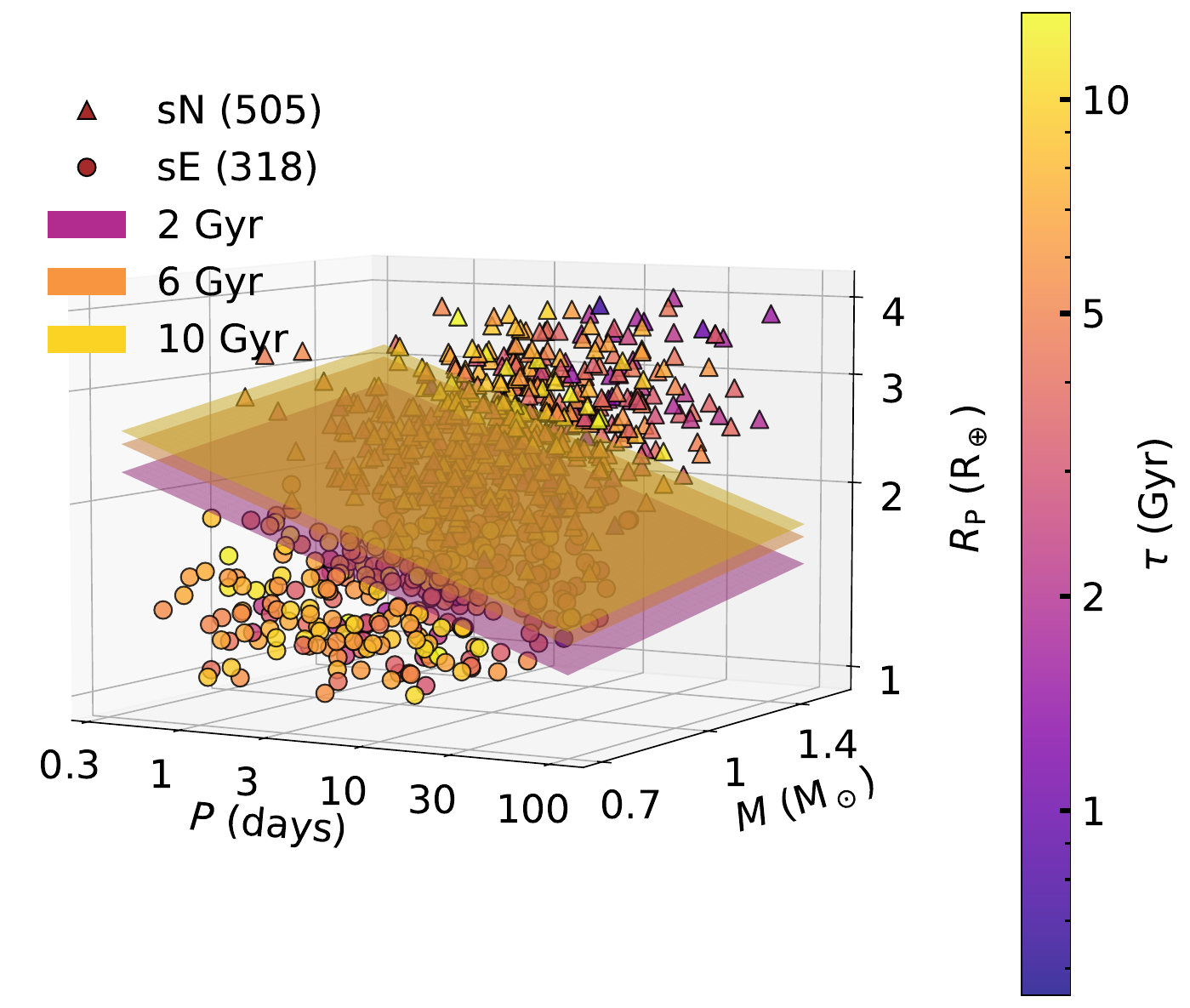}
            \caption{Four-dimensional exoplanet radius valley across orbital period, stellar mass, and age space. Observed planets are overlaid, colored by age, with sub-Neptunes (sN) and super-Earths (sE) distinguished by marker type. The fitted planes show the valley at representative ages of 2, 6, and 10 Gyrs.}
            \label{4D_valley}
\end{figure}

We find coefficients of $\alpha = -0.08^{+0.02}_{-0.02}$, $\beta = 0.31^{+0.09}_{-0.08}$, and $\gamma = 0.07^{+0.03}_{-0.04}$, and intercept of $\delta = 0.30^{+0.04}_{-0.03}$. In Fig.~\ref{4D_valley}, we present a three-dimensional visualisation of the radius valley in the space of planet radius, orbital period, and stellar mass, with points colour-coded according to stellar age. The derived coefficients in four-dimensional fit are consistent with the corresponding slopes in two dimensions based on the gapfit method ($-0.12^{+0.02}_{-0.01}$ in orbital period, $0.19^{+0.09}_{-0.07}$ in stellar mass, and $0.00^{+0.05}_{-0.03}$ in stellar age). They are also consistent with the four-dimensional coefficients reported by \citet{ho2023deep}, derived from a linear support vector machine fit to orbital period, stellar mass, and stellar age. The slightly higher stellar age slope in our analysis, $0.07^{+0.03}_{-0.04}$ vs.\ their $0.03^{+0.02}_{-0.03}$ \citep{ho2023deep}, likely reflects differences in stellar age estimates arising from variations in the underlying physics of the stellar models (see Fig.~\ref{MAISTEP_F18_1} in the Appendix.)  

Our four-dimensional radius valley provides strong support for core-powered mass-loss models, as the evolution unfolds on Gyr timescales \citep[][]{ginzburg2016super,ginzburg2018core,gupta2019sculpting,gupta2020signatures}. On the other hand, photoevaporation is predicted to be most effective within the first $\sim 100$ Myr when high-energy output from Sun-like stars peaks \citep[e.g.,][]{jackson2012coronal,tu2015extreme}, although some fraction of sub-Neptunes are predicted to transform to super-Earths on longer timescales \citep[][]{rogers2021photoevaporation}. Moreover, \citet{king2021euv} found that UV radiation remains substantial beyond 100 Myr, suggesting that the process may extend longer than previously thought. We also note, however, that age trends remain inconclusive because of substantial uncertainties in stellar age estimates; accordingly, these interpretations should be regarded with caution.

\section{Summary and conclusions}
\label{conclusion}

In this study, we analysed a sample of 779 stars from the SWEET-Cat database, hosting 893 confirmed planets listed in the NASA Exoplanet Archive, to revisit the size distribution of small, close-in exoplanets. Our primary focus was to re-examine the existence and depth of the radius valley, and how its location varies with orbital period, incident flux, stellar mass, and stellar age. We derived stellar radii, masses, and ages using MAISTEP, a grid-based machine-learning stellar parameter estimator, achieving median fractional uncertainties of 2\%, 2\%, and 27\%, respectively, using only effective temperatures and metallicities from spectroscopy, along with $Gaia$-based luminosities. We note that we are able to achieve the radius and mass precision typically expected when asteroseismic data are available. Planetary radii were updated accordingly, reaching a median fractional uncertainty of 4\%.
 
Our revised planet radii reveal a well-defined bimodal distribution, with peaks near $\sim$1.5 R$_\oplus$ (super-Earths) and $\sim$2.6 R$_\oplus$ (sub-Neptunes), separated by a valley around 2 R$_\oplus$, consistent with previous studies. We find that the radius valley is partially filled and deeper than in the distribution based on the planetary radii in the NASA Exoplanet Archive (Fig.~\ref{pl_radius_distributions}).

The location of the valley shifts toward smaller radii at longer orbital periods (i.e., lower incident flux), yielding a power-law slope of $-0.12^{+0.02}_{-0.01}$ when modelled in terms of orbital period, or equivalently $0.10^{+0.02}_{-0.03}$ with respect to incident flux (Fig.~\ref{pl_rade_orbper} \& Fig.~\ref{pl_rade_inc_flux}, respectively). 

The average size of planets within the radius valley 
increases with host star's mass, with a slope of $0.19^{+0.09}_{-0.07}$. The characteristic sizes of sub-Neptunes and super-Earths show trends with slopes of $0.17^{+0.04}_{-0.04}$ and $0.11^{+0.05}_{-0.05}$, respectively, indicating a stronger dependence in sub-Neptunes than in super-Earths (Fig.~\ref{pl_rade_star_mass}). 

Regarding stellar age, we find that the ratio of super-Earths to sub-Neptunes increases from $0.51^{+0.11}_{-0.08}$ in systems younger than 3 Gyr to $0.64^{+0.12}_{-0.11}$ in older systems (Fig.~\ref{pl_rade_3Gyr_age_bins}). The radius valley also becomes shallower and shifts to larger radii in the older systems, confirming age-dependent evolution in planet demographics over gigayear timescales. A four-dimensional (planetary radius, orbital period, stellar mass, and stellar age) characterisation of the valley yields slopes in orbital period and mass similar to those in two dimensions (planet radius-orbital period and planet radius-stellar mass), and a slope of $0.07^{+0.03}_{-0.04}$ in stellar age. The observed Gyr-scale trends are consistent with core-powered mass loss as an important driver of atmospheric escape, although photoevaporation cannot be fully ruled out, given theoretical expectations that it may continue to operate on timescales exceeding $\sim 100$ Myrs.   

Finally, the average size of sub-Neptunes does not show a decline with time. This lack of an observed decline may stem from the large uncertainties in our host age estimates, despite theoretical expectations that sub-Neptunes gradually shrink and evolve into super-Earths. These findings underscore the need for more precise host star age measurements, which future asteroseismic observations, such as those expected from ESA’s $PLATO$ mission \citep[][]{rauer2014plato}, are likely to provide.

\begin{acknowledgements} 
\footnotesize
We thank the referee for the insightful suggestions that improved the manuscript. J.K. acknowledges funding through the Max-Planck Partnership group - SEISMIC between Max-Planck-Institut für Astrophysik (MPA) - Germany and Kyambogo University (KyU) - Uganda. 
B.N. and O.T. acknowledge funding through the Max Planck - Humboldt research unit established in Uganda at Kyambogo University. B.N. also acknowledges funding from the UNESCO-TWAS programme, ``Seed Grant for African Principal Investigators'' financed by the German Ministry of Education and Research (BMBF). T.L.C is supported by Funda\c c\~ao para a Ci\^encia e a Tecnologia (FCT) in the form of a work contract (\href{https://doi.org/10.54499/2023.08117.CEECIND/CP2839/CT0004}{2023.08117.CEECIND/CP2839/CT0004}). S.G.S acknowledges the support from FCT through Investigador FCT contract nr. CEECIND/00826/2018 and  POPH/FSE (EC). N.M. acknowledges the research theory grant ''Synergic tools for characterizing solar-like stars and habitability conditions of exoplanets"  under the INAF national call for Fundamental Research 2023. V.A. and N.C.S. acknowledge support from FCT – Funda\c{c}\~ao para a Ci\^encia e Tecnologia through national funds and from FEDER via COMPETE2020 – Programa Operacional Competitividade e Internacionalização, under the grants UIDB/04434/2020 (DOI: 10.54499/UIDB/04434/2020) and UIDP/04434/2020 (DOI: 10.54499/UIDP/04434/2020). V.A. is also supported through a work contract funded by the FCT Scientific Employment Stimulus program (reference 2023.06055.CEECIND/CP2839/CT0005, DOI: 10.54499/2023.06055.CEECIND/CP2839/CT0005). Views and opinions expressed are however those of the author(s) only and do not necessarily reflect those of the European Union or the European Research Council. Neither the European Union nor the granting authority can be held responsible for them. 

$Software$: Data analysis in this manuscript was carried out using the Python  3.10.12 libraries; scikit-learn 1.5.1 \citep{pedregosa2011scikit}, pandas 2.2.2 \citep{mckinney2010proceedings}, NumPy 1.26.4 \citep{van2011numpy}, and SciPy 1.14.1 \citep{virtanen2020scipy}. 
 \end{acknowledgements}



\bibliographystyle{aa}
\bibliography{mybib} 



\appendix
\section{Comparison of stellar parameters}
\label{appendix}
Exploration of the radius valley demographics is critical to the accuracy and precision of the stellar radius, mass, and age. In \citet{kamulali2025maistep}, the stellar parameters inferred from MAISTEP were validated. Here, we compare MAISTEP-inferred parameters  with the literature values for the subset of stars in common. First, we compare stellar parameters from MAISTEP for 1,221 stars to those derived using a Bayesian and Markov Chain Monte Carlo–based tool, Systematic Inference on a Massive Scale \citep[SPInS;][]{lebreton2020spins}. Both methods incorporate the same observational constraints: $T_{\text{eff}}$, [Fe/H], and $L$, and rely on the same grid of stellar evolutionary models. We find excellent agreement in the inferred radius, with a median offset of  0.3\% and a scatter of 1.1\%. The median fractional uncertainty from both methods is 2\%. For stellar mass, we obtain a median offset of 0.2\% and a scatter of 2.6\%. The fractional uncertainties are 2\% for MAISTEP and 7\% for SPInS. The offset and scatter in stellar ages are 0.6\% and 13.1\%, respectively. We observe higher age estimates for some stars with SPInS-inferred ages between 6 and 11 Gyr. The fractional uncertainty in ages is 27\% for MAISTEP, roughly half of the 47\% recorded from SPInS. The observed differences in uncertainties may reflect the distinct ways the two routines handle uncertainties in the input parameters.  

\begin{figure}[t]
        \centering
        \includegraphics[scale=0.3]{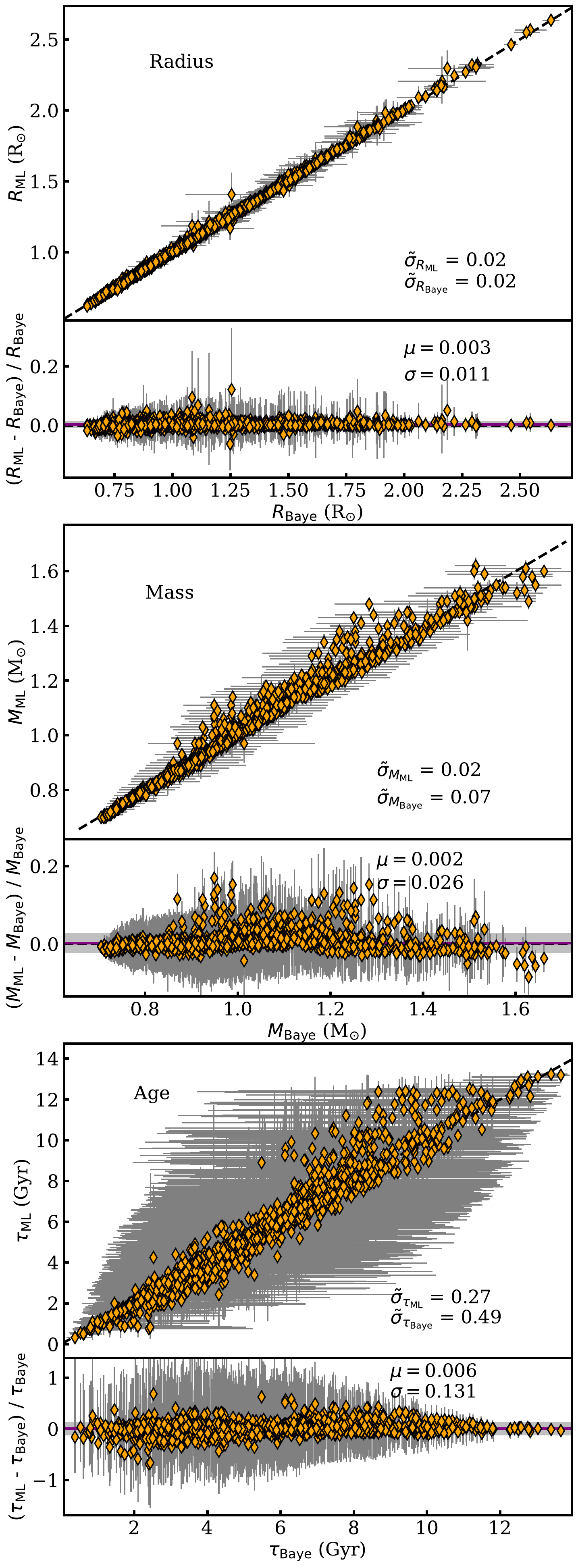}
            \caption{One-to-one relation and fractional differences from MAISTEP (ML) with respect to those from SPInS (Baye) in radius (top), mass (middle), and age (bottom).
            The black dashed lines in all panels represent the unity relation.
            The purple line in the fractional difference plots represents the median offset, $\mu$. The gray horizontal bands highlights the associated scatter, $\pm\sigma$. We also indicate the median fractional uncertainties in the one-to-one plots using $\tilde{\sigma}$.}
            \label{MAISTEP_SPInS}
\end{figure}

\begin{figure}[t]
        \centering
        \includegraphics[scale=0.3]{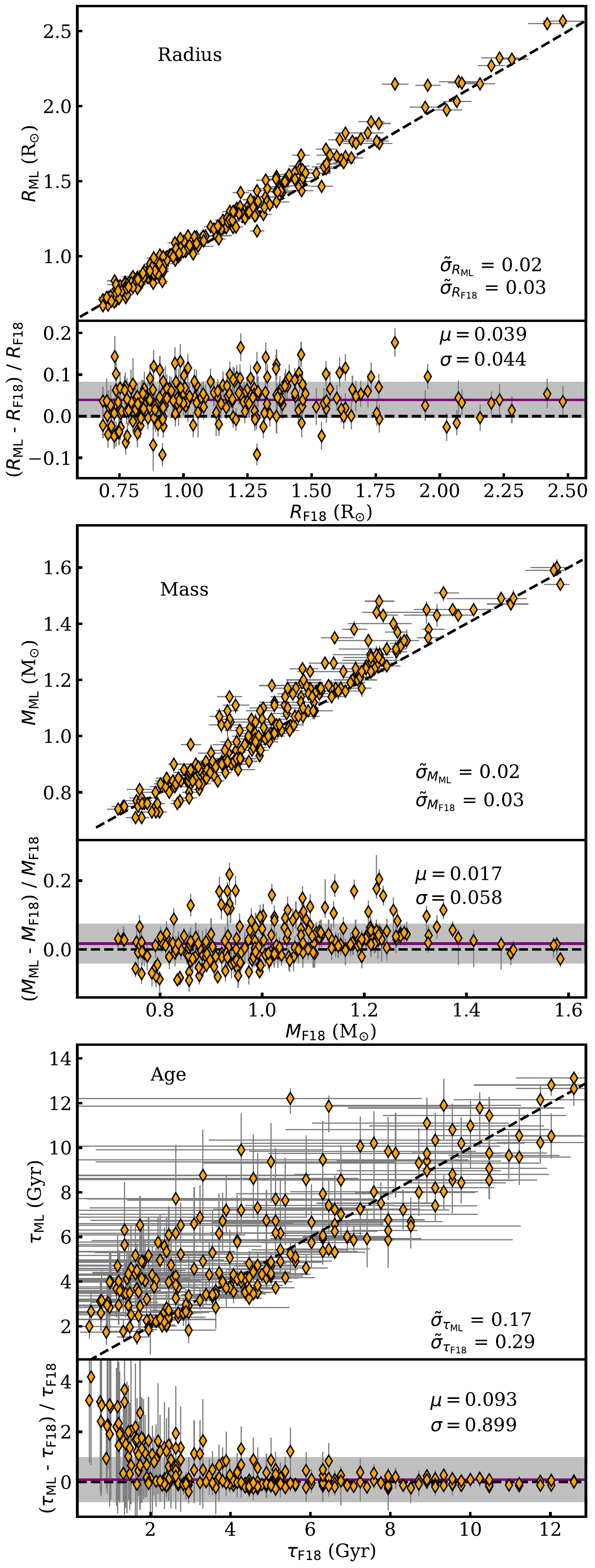}
            \caption{Same as Fig.~\ref{MAISTEP_SPInS}, but showing MAISTEP inferences using zero-point parallax corrections that depend on each star’s magnitude, color, and position in the luminosity constraint, compared with the estimates from \citet{fulton2018california} for an overlapping sample of 236 stars.}
            \label{MAISTEP_F18_1}
\end{figure}

\begin{figure}[t]
        \centering
        \includegraphics[scale=0.3]{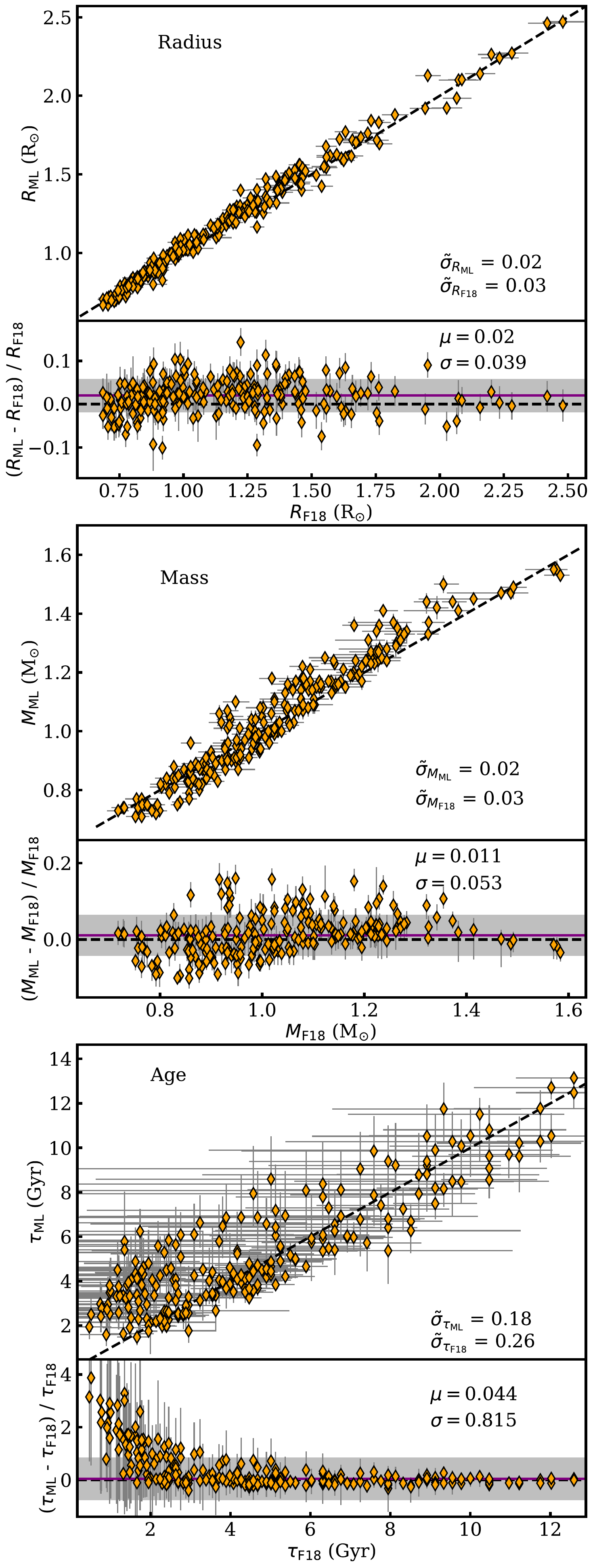}
            \caption{Same as Fig.~\ref{MAISTEP_F18_1}, but showing MAISTEP inferences using a fixed -0.053 zero-point parallax correction in the luminosity constraint, compared with the estimates from \citet{fulton2018california} for an overlapping sample of 236 stars.}
            \label{MAISTEP_F18_2}
\end{figure}

Next, we compare with the homogeneous estimates of \citet{fulton2018california} for an overlapping sample of 226 stars, shown in Fig.~\ref{MAISTEP_F18_1}. We note that this sample has uniform spectroscopic constraints from \citet{petigura2017california}, which were used as input for the stellar parameter determinations in \citet{fulton2018california}. Consequently, nearly all MAISTEP inferences for this subset rely on the same uniform spectroscopic dataset, ensuring consistency with \citet{fulton2018california}.  
The radii, masses, and ages from MAISTEP are systematically higher by 3.9\%, 1.7\%, and 9.3\%, with scatters of 4.4\%, 5.8\%, and 90\%. The offsets arise partly from the fixed parallax zero-point correction of -0.053 adopted in \citet{fulton2018california} for the $Gaia$ DR2 data, which impacts distance estimates and consequently the luminosity values. As discussed in the main text, we use $Gaia$ DR3 parallaxes and $G$-band magnitudes, with zero-point parallax corrections that depend on the star's magnitude, color, and position \citep[][]{lindegren2021gaia}. Our derived zero-point correction have a median value of -0.027 mas, about half of -0.053 mas. Adopting a fixed -0.053 mas in our luminosity calculations reduces the offsets to 2\% in radius, 1.1\% in mass, and 4.4\% in age (Fig.~\ref{MAISTEP_F18_2}). For a small subset of stars, slightly higher masses are apparent, a feature seen in MAISTEP mass predictions in comparison to those from SPInS (Fig.~\ref{MAISTEP_SPInS}). In addition, the stellar parameters in \citet{fulton2018california} were derived based on the MESA Isochrones and Stellar Tracks \citep[][]{choi2016mesa,dotter2016mesa} via \texttt{isoclassify} \citep{huber2017asteroseismology}, a Bayesian stellar pipeline, incorporating spectroscopic $T_{\text{eff}}$ and [Fe/H] from  \citep{petigura2017california}, along with $Gaia$ DR2 parallaxes and 2MASS $K$-band photometry \citep[][]{skrutskie2006two}.
In \texttt{isoclassify}, distances are inferred from parallax using a Bayesian approach that adopts an exponentially decreasing volume-density prior with a 1.35 kpc scale length \citep{astraatmadja2016estimating}, rather than taking the inverse parallax as in equation~\ref{lum_eqn}. This difference may contribute to the remaining 2\% radius offset. The observed discrepancies in stellar ages are most likely driven by differences in the underlying stellar-evolution model physics. This is based on the fact that MAISTEP and SPInS give consistent results when run on the same stellar-model grid and constraints, with the exception of a few stars with ages $\gtrsim 6$ Gyr (see Fig.~\ref{MAISTEP_SPInS}). 

\begin{figure}[t]
        \centering
        \includegraphics[scale=0.3]{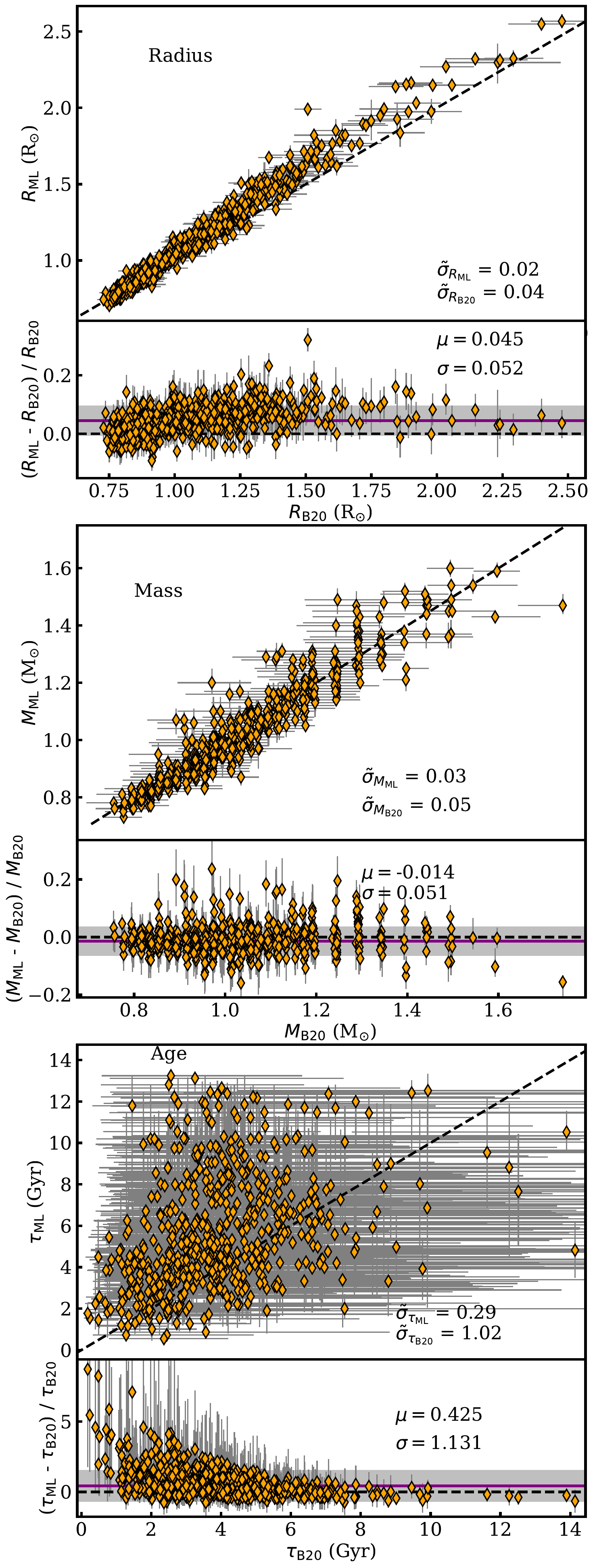}
            \caption{Same as Fig.~\ref{MAISTEP_F18_1}, but comparing to isochrone estimates of 587 stars in \citet{2023arXiv230111338B}.}
            \label{MAISTEP_B20}
\end{figure}

We also compare MAISTEP inferences with those reported in \citet{2023arXiv230111338B} for a common sample of 587 stars. Their stellar parameters were derived using isochrone models similar to \citet{fulton2018california} and the \texttt{isoclassify} tool. As shown in Fig.~\ref{MAISTEP_B20}, the MAISTEP estimates are systematically higher and consistent with earlier comparisons. \citet{2023arXiv230111338B} used Gaia DR3 $G_{BP}$ and $G_{RP}$
 magnitudes, parallaxes (with a zero-point correction following the same procedure we adopt), and metallicities as constraints. We therefore attribute the differences, at least in part, to the absence of $T_{\text{eff}}$
 from their constraint set.

\begin{figure}[t]
        \centering
        \includegraphics[scale=0.3]{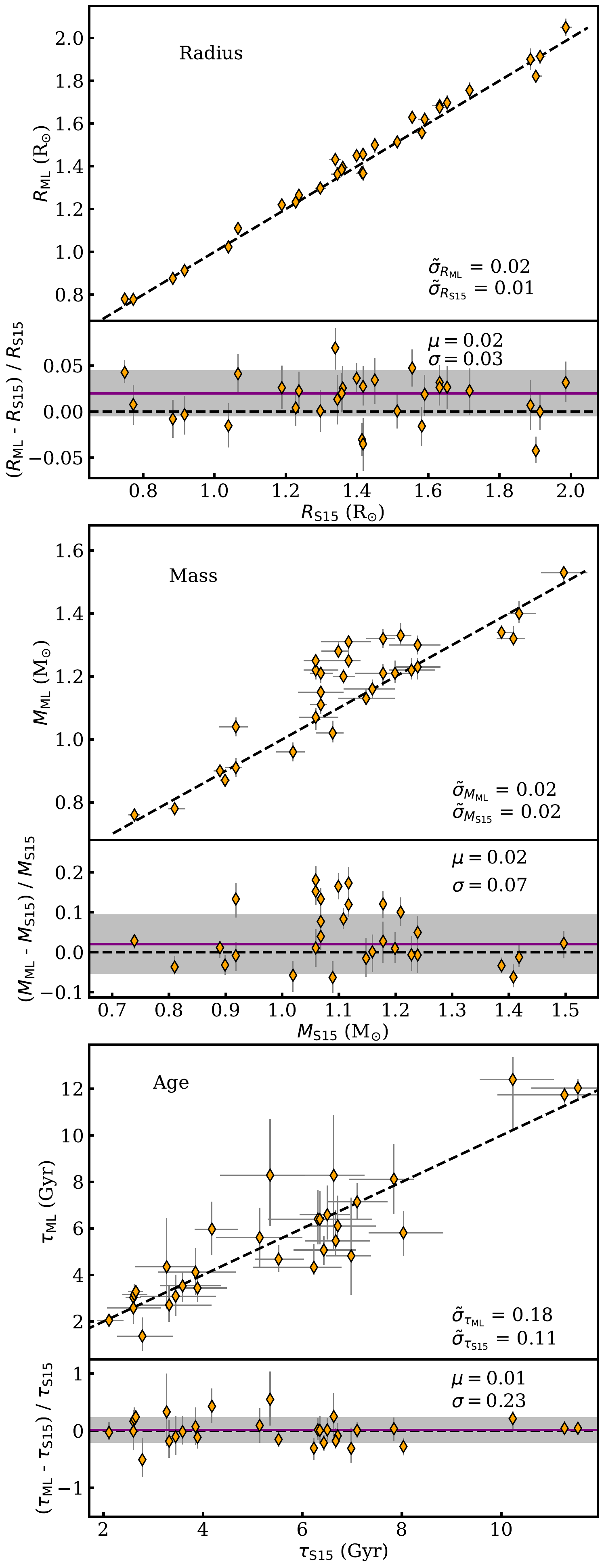}
            \caption{Same as Fig.~\ref{MAISTEP_F18_1}, but comparing to seismic estimates for 31 exoplanet host stars in \citet{silva2015ages}.}
            \label{MAISTEP_S15}
\end{figure}

Lastly, we apply MAISTEP to 31 seismic planet-host stars from \citet{silva2015ages}, excluding two stars, which are likely  binary components with RUWE > 1.4 (Fig.~\ref{MAISTEP_S15}). We use their compiled spectroscopic constraints, along with luminosities from equation~\ref{lum_eqn}; only five stars overlap with our main sample. Offsets (and scatter) of 1\% (3\%) in radius, 2\% (7\%) in mass, and 1\% (23\%) in age are obtained, demonstrating a good agreement.

\label{lastpage}
\end{document}